\documentclass[10pt,twocolumn, prl]{revtex4} 

\usepackage{amsmath}
\usepackage{amssymb}
\usepackage{graphicx}
\renewcommand{\small}[1][]{#1}
\usepackage{upgreek}
\usepackage{booktabs}  
\AtBeginDocument{
\heavyrulewidth=.08em
\lightrulewidth=.05em
\cmidrulewidth=.03em
\belowrulesep=.65ex
\belowbottomsep=0pt
\aboverulesep=.4ex
\abovetopsep=0pt
\cmidrulesep=\doublerulesep
\cmidrulekern=.5em
\defaultaddspace=.5em
}

\usepackage{bbold}  

\usepackage{lipsum}
\usepackage{color}
\usepackage[dvipsnames]{xcolor}

\newcommand{\hide}[1]{\relax}
\newcommand{\unit}[1]{\ensuremath{\,\mathrm{#1}}}
\newcommand{\Og}{\ensuremath{\Omega}}

\newcommand{\Om}{\ensuremath{\Omega_\mathrm{m}}}

\newcommand{\Gm}{\ensuremath{\Gamma_\mathrm{m}}}

\usepackage{array}

\newcommand{\Geff}{\ensuremath{\Gamma_\mathrm{eff}}}
\newcommand{\Gmeas}{\ensuremath{\Gamma_\mathrm{opt}}}
\newcommand{\Gdec}{\ensuremath{n \Gamma_\mathrm{m}}}

\newcommand{\oL}{\ensuremath{\omega_\mathrm{l}}}

\newcommand{\vcr}{\ensuremath{g_0}}
\newcommand{\ecr}{\ensuremath{g}}

\newcommand{\kT}{\ensuremath{\kappa_\mathrm{T}}}

\newcommand{\ncav}{\ensuremath{\bar{n}_\text{cav}}}
\newcommand{\nth}{\ensuremath{n_\text{th}}}
\newcommand{\mean}[1]{\langle#1\rangle}

\newcommand{\chim}{\ensuremath{\chi_\mathrm{m}}}
\newcommand{\chieff}{\ensuremath{\chi_\mathrm{eff}}}

\newcommand{\dd}{\mathrm{d}}
\newcommand{\e}{\mathrm{e}}

\newcommand{\etaDet}{\eta_\mathrm{d}}

\AtEndDocument{\clearpage}

\begin{document}

\title{Multimode optomechanical system in the quantum regime}
\author{W.\ H.\ P.\ Nielsen$^{1}$, Y.\ Tsaturyan$^{1}$,  C.\ B.\ M{\o}ller $^{1}$, E.\ S.\ Polzik$^{1}$, A. Schliesser$^{1}$}
\email{albert.schliesser@nbi.dk}
\affiliation{$^{1}$Niels Bohr Institute, University of Copenhagen, 2100 Copenhagen, Denmark}

\begin{abstract}
We realise a simple and robust optomechanical system with a multitude of long-lived ($Q>10^7$) mechanical modes in a phononic-bandgap shielded membrane resonator. 
%
An optical mode of a compact Fabry-Perot resonator detects these modes' motion with a measurement rate ($96\unit{kHz}$) that exceeds the mechanical decoherence rates already at moderate cryogenic temperatures ($10\,\mathrm{K}$).
Reaching this quantum regime entails, i.~a., quantum measurement backaction exceeding thermal forces, and thus detectable optomechanical quantum correlations.
In particular, we observe ponderomotive squeezing of the output light mediated by a multitude of mechanical resonator modes, with quantum noise suppression up to -2.4~dB (-3.6~dB if corrected for detection losses) and bandwidths $\lesssim 90\,\mathrm{ kHz}$.
%
%
The multi-mode nature of the employed membrane and Fabry-Perot resonators lends itself to hybrid entanglement schemes involving multiple electromagnetic, mechanical, and spin degrees of freedom.
\end{abstract}

\maketitle

Within the framework of quantum measurement theory \cite{Braginsky1992, Clerk2010}, quantum backaction (QBA) enforces Heisenberg's uncertainty principle:
it implies that any `meter' measuring a system's physical variable induces random perturbations on the conjugate variable.
Optomechanical transducers of mechanical motion \cite{Braginsky1992, Clerk2010,Aspelmeyer2014} implement weak, linear measurements, whose QBA is typically small compared to thermal fluctuations in the device.
Nonetheless, recent experiments have evidenced QBA in continuous position measurements of mesoscopic (mass $m\lesssim 10 \unit{ng}$) mechanical oscillators \cite{Murch2008, Purdy2013, Teufel2016}.
While QBA appears as a heating mechanism \cite{Teufel2016,Peterson2016} from the point of view of the mechanics only, it  correlates the fluctuations of mechanical position with the optical meter's quantum noise. 
These correlations are of fundamental, but also practical interest, e.~g.~as a source of entanglement  and a means to achieve measurement sensitivities beyond standard quantum limits \cite{Vitali2007, Genes2008a, Khalili2012, Buchmann2016}.
Correspondingly, they have been much sought-after experimentally \cite{Verlot2009, Brooks2012, Safavi-Naeini2012, Purdy2013a,Safavi-Naeini2013, Weinstein2014, Underwood2015, Sudhir2016}.
Quantum correlations in \emph{multimode} systems supporting many mechanical modes give rise to even richer physics and new measurement strategies \cite{Mancini2002, Hartmann2008, Bhattacharya2008b, Houhou2015, Woolley2013, Buchmann2015}.
However, while quantum electromechanical coupling to several mechanical modes has been studied \cite{Massel2012, Noguchi2016}, quantum fluctuations have so far only been investigated for a pair of collective motional modes of $\sim900$ cold atoms trapped in an optical resonator \cite{Spethmann2016}.
In contrast, QBA cancellation and entanglement have been extensively studied with atomic spin oscillators \cite{Hammerer2010,Krauter2011,Vasilakis2015}.

\begin{figure}[thb]
\includegraphics[width= .85\linewidth]{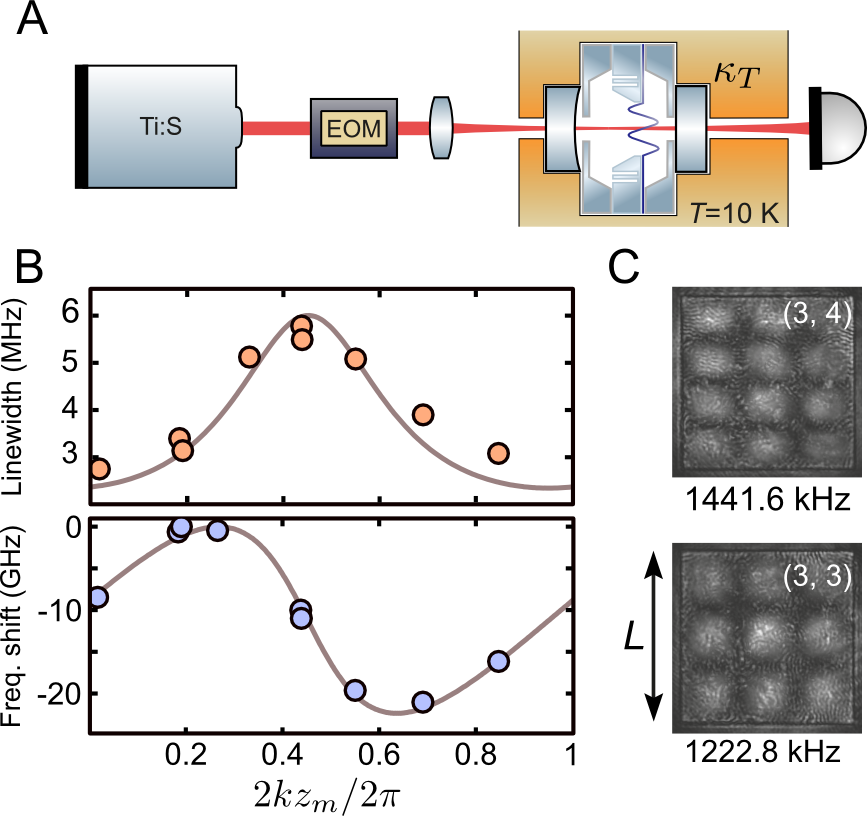}
\caption{
Multimode optomechanical system. 
A: Optical setup, in which a low-noise Ti:S laser with modulation sidebands from an electro-optic modulator (EOM) pumps a Fabry-Perot resonator held in a cryostat. The resonator contains the sample chip with the nanomechanical membrane and two spacer chips. 
B: Turning of optical resonator linewidth and frequency with membrane position with respect to the wavelength. 
C: Dark-field images of two mechanical modes.
}
\label{fig1}
\end{figure}

In our study, we use highly stressed, $\sim60\unit{nm}$ thick  Si$_{3}$N$_{4}$ nanomechanical membrane resonators.
They naturally constitute a multimode systems, supporting  mechanical modes at frequencies $\Omega_\mathrm{m}^{(i,j)}=\Omega_\mathrm{m}^{(1,1)}\sqrt{(i^2+j^2)/2}$ in the MHz range, of which two examples are shown in Fig.~\ref{fig1}C.
The membrane is embedded in a 1.7~mm long Fabry-Perot resonator held at a temperature $T\approx10\unit{K}$ in a  simple flow cryostat (Fig.~\ref{fig1}).
The location $z_\mathrm{m}$ of the membrane along the standing optical waves (wavelength $2\pi/k$) then determines an optical frequency shift $\Delta f_\mathrm{cav}$, as well as the resonance linewidth $\kappa$ \cite{Jayich2008, Wilson2009, SI}.
As an optimal working point we choose $2kz_m/2\pi\approx 0.43$ where the optomechanical coupling  $G/2\pi=\partial{f_\text{cav}}/\partial{z_m}$ is largest (Fig.\ \ref{fig1}C), and the biggest fraction $\kT/\kappa$ of scattered photons exits the resonator through the `transmission' port towards the detector \cite{SI}.

One key challenge in the generation and observation of optomechanical quantum correlations is thermal decoherence of the mechanics, which occurs at a rate $\Gdec=k_\mathrm{B} T/\hbar Q$.
Here, $n$ is the mode occupation in equilibrium with the bath of temperature $T\approx 10 \unit{K}$, while $\Gm$ is the mechanical dissipation rate and $Q=\Om/\Gm$ (dropping mode indices $(i,j)$ for convenience).
For the multimode system studied here, this necessitates ultra-high mechanical $Q$-factors across a wide frequency range, which we achieve via a phononic bandgap shield.
By embedding the membrane in a periodically patterned silicon frame, we suppress phonon tunnelling loss into elastic modes of the substrate, thereby consistently enabling ultra-low mechanical dissipation \cite{Tsaturyan2014, SI}.

\begin{figure}[tb]
\includegraphics[width=1\linewidth]{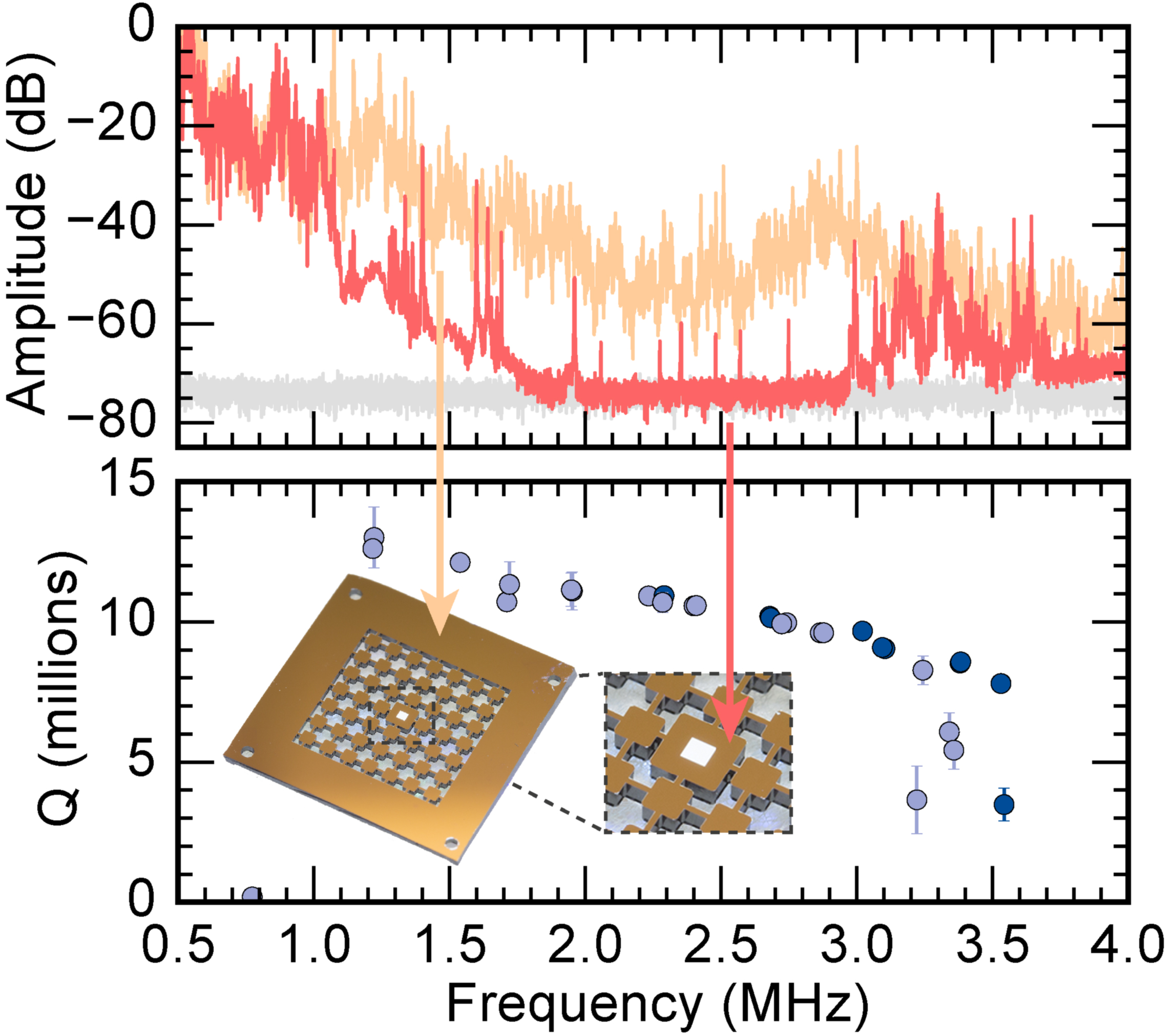}
\caption{
{A: Response of the sample frame (orange) and the sample centre containing the membrane (red)  to an acoustic excitation of the sample frame, showing broadband suppression of phonon propagation down to the measurement background (grey). 
B: Resulting membrane mode $Q$-factors (light and dark blue circles), showing consistently $Q\gtrsim10^7$ in the protected frequency region---also for low index modes with $i\vee j < 3$ (light blue)---but not outside.
Inset shows photograph of the actual sample.}
}
\label{fig:bandgapQs}
\end{figure}

To characterise the degree of acoustic isolation achieved, a prototype chip with a membrane of side-length $L=547\unit{\upmu m}$ is mounted on a swept-frequency piezo shaker.
Under this excitation, the phononic `defect' 
that hosts the membrane in the center of the shield moves about $20\unit{dB}$ less
than the sample's outer frame, see Fig.\ \ref{fig:bandgapQs}A.
While this experiment probes the suppression of a subset  of elastic modes only, we emphasise that the used shield provides a full phononic bandgap, i.\ e.\ no modes exist in this frequency region \cite{Tsaturyan2014, SI}. 
Furthermore, the small size of the defect ($\sim1.3\unit{mm}$) in direct contact with the membrane results in a sparse background phononic density of states (see Fig.\ \ref{fig:bandgapQs}B), entailing a low number of  membrane-defect hybrid modes.

Figure \ref{fig:bandgapQs}B shows the effect on the $Q$-factor of the 30 lowest-frequency mechanical modes.
Clearly, the values scatter for modes outside the shielded $~1$-$3 \unit{MHz}$ region, while all modes in the bandgap achieve $Q\gtrsim10^7$.
Importantly, this holds also for low-index modes with $i$ or $j < 3$, rendering our observations consistent with the full elimination of dissipation by elastic wave radiation \cite{Wilson-Rae2011,Villanueva2014}.

Returning to the membrane-in-the-middle  system of Fig.~\ref{fig1}B, we note that the optical mode width on the membrane is sufficiently small ($w= 39\unit{\upmu m}$) to resolve all relevant mechanical mode patterns.
The vacuum optomechanical coupling rates are then determined by the modes' displacement at the location $(x,y)$ of the optical beam in the membrane plane \cite{SI}
\begin{equation}
  g_0^{(i,j)}(x,y)\approx  G  \cdot x_\text{ZPF}^{(i,j)}\hspace{1pt} 
 \sin\left(\frac{\pi i x}{L_i}\right) \sin\left(\frac{\pi j y}{L_j}\right),
  \label{eq:g0s}
\end{equation}
where $x_\text{ZPF}^{(i,j)}=\sqrt{\hbar/2 m \Omega_\mathrm{m}^{(i,j)} }$ is the mechanical zero-point fluctuation amplitude.

\begin{figure*}[tb]
\includegraphics[width=1\linewidth]{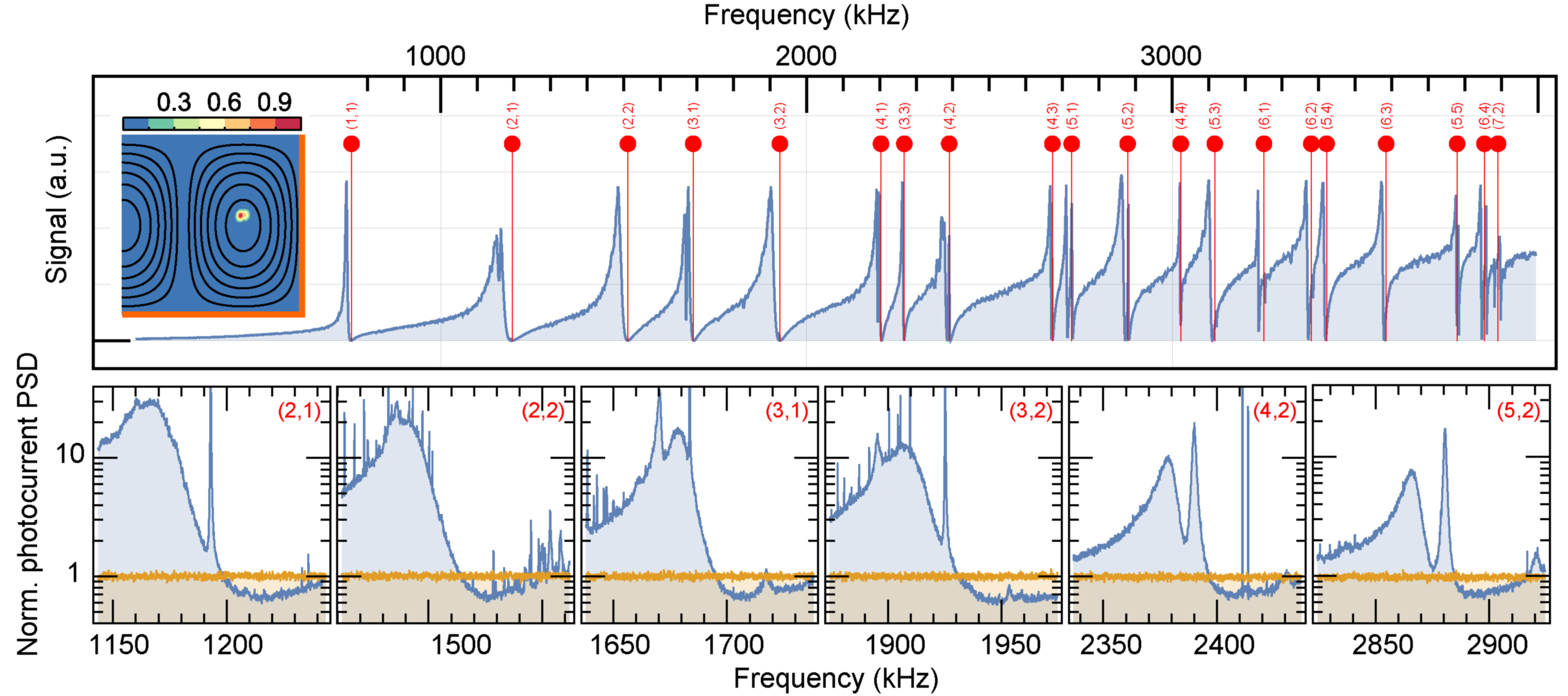}
\caption{
Top: Multimode optomechanically induced transparency (OMIT) in the cavity response, with expected frequencies $\Omega_\mathrm{m}^{(i,j)}$ indicated by red lines, labeled with the mode index.
Inset shows the extracted location of the optical beam within one quadrant of the membrane, with color-coded normalised probability density. 
Contours of equal displacement of the (3,2) mode are also shown, with the membrane's  clamped edges indicated in orange.
B: Strong simultaneous light squeezing from six mechanical modes. Blue traces are the recorded cavity output spectra, yellow is the shot noise level.
}
\label{fig:broaddataoverview}
\end{figure*}

To extract these rates for a membrane with $L\approx544\unit{\upmu m}$ 
and $m=62$ ng, we probe the weakly driven optical resonator (linewidth $\kappa/2\pi=14\unit{MHz}$ at $2\pi/k=799.877\unit{nm}$) with an additional optical sideband generated by an electro-optic modulator (EOM).
A broad frequency scan reveals Optomechanically Induced Transparency (OMIT \cite{Weis2010}) features for more than 30 modes, as shown in Fig.~\ref{fig:broaddataoverview}A.

The extracted vacuum coupling rates differ widely for different modes, and range up to $\sim115\unit{Hz}$.
The broadband `fingerprint' spectrum reveals, in addition, the mechanical mode frequencies, whose $i$-$j$ degeneracies appear all lifted with $L_i\approx 0{.}994 L_j$, in reasonable agreement with a $0{.}4\%$ difference in membrane side lengths measured in a microscope image. 
This lifted degeneracy motivates the assumption that membrane-membrane mode hybridisation \cite{Chakram2013} is negligible in this device.
While this assumption is not critical for the main conclusions of this work, it allows a simple inversion of the relations (\ref{eq:g0s}) to localise the optical beam position on the membrane (Fig.~\ref{fig:broaddataoverview}A).

To realise optomechanical quantum correlations, QBA---here essentially the quantum fluctuations of radiation pressure on the membrane---must exceed the thermal Langevin force. 
In the unresolved sideband case $\Om\ll\kappa$ considered here, this translates to $1<\bar S_{FF}^\mathrm{qba}(\Og)/\bar S_{FF}^\mathrm{th}(\Og)\approx \Gmeas/\Gdec\equiv C_\mathrm{q}$, where $ \Gmeas=4 \vcr^2 \ncav/\kappa$
is the optomechanical measurement rate \cite{Clerk2010}, $\ncav$ the average number of intracavity photons, and $C_\mathrm{q}$ the quantum cooperativity.
Remarkably, due to the consistently ultrahigh $Q$-factors, this condition can be fulfilled for a multitude of mechanical modes, even at $T=10\unit{K}$, in the system reported here.

To evidence continuous variable quantum correlations, and realise quantum-limited measurements in general, high detection efficiency is a second requirement---lest entangled meter states are replaced by ordinary vacuum. 
In contrast to both microwave and optical experiments that deploy  advanced cryogenic  technologies \cite{Palomaki2013a, Weinstein2014, Verhagen2012, Wilson2015, Peterson2016}, the simplicity of our setup (Fig.~\ref{fig1}B) affords a high detection efficiency $\etaDet=80\%$.
Combined with a largely one-sided cavity, the probability for an intracavity sideband photon to be recorded as a photoelectron is expected to be $\eta=\etaDet \kT/\kappa=77\%$.

Ponderomotive squeezing \cite{Fabre1994, Mancini1994} provides one of the most model-agnostic and straightforward-to-calibrate manners to gauge the presence or absence of optomechanical quantum correlations. 
It originates from the correlations that radiation pressure creates between the quantum fluctuations of the light's amplitude quadrature $X$ and the membrane position $q$.
As the latter, in turn, shifts the phase $Y$ of the intracavity field, amplitude-phase quantum correlations in this field are created.

 A slightly detuned cavity ($|\Delta|\ll\kappa$) rotates the optical quadratures so that the quantum correlations  appear as sub-vacuum noise in the output light amplitude $X_\mathrm{out}$ \cite{SI}. 
Figure~\ref{fig:broaddataoverview}B shows the measured spectrum  $\bar S_{XX}^\mathrm{meas}(\Omega)$ of this entity, after propagation to the detector.
Here, the driving laser is held at the detuning $\Delta/2\pi=-1.8\unit{MHz}$ of the OMIT measurement, but the EOM is deactivated. 
Depressions in the noise level appear close to the eigenfrequencies of strongly coupled modes, of which six are shown.
A comparison with an independent measurement of optical vacuum noise \cite{SI} reveals significant ponderomotive squeezing in all these spectral regions.
The maximum observed squeezing is observed around the $(3,2)$ mode, and amounts to $-2.4\unit{dB}$ (or $-3.6\unit{ dB } $ if corrected for detection losses $\etaDet$) exceeding all previously reported values \cite{Brooks2012,Purdy2013a,Safavi-Naeini2013}.

For a quantitative discussion of these results, we invoke a description of the system using a Heisenberg-Langevin approach.
The output amplitude fluctuation spectrum of an ideal system can be calculated using a covariance matrix approach \cite{Fabre1994, Mancini1994}, and simplified to the  intuitive 
\begin{align}
	\bar S_{XX}^\mathrm{out}(\Omega)&\approx
		1-2\frac{8 \Delta}{\kappa}\Gmeas\mathrm{Re}\left\{\chieff(\Og) \right\}+\\
		&+\left(\frac{8 \Delta}{\kappa}\right)^2 \Gmeas \left|\chieff(\Og)\right|^2\nonumber
		\left(\Gmeas +n \Gm \right)
\end{align}
for the high-cooperativity, non-resolving cavity ($4 \vcr^2 \ncav/\Gm\gg\kappa\gg\Om,\Delta$) considered here \cite{SI}.
Note that $\chieff(\Og)$ is the effective mechanical susceptibility taking into account the dynamical backaction (cooling) of the detuned laser \cite{Schliesser2006, Arcizet2006a}.
If the correlation term ($\propto \mathrm{Re}\left\{\chieff(\Og) \right\}$) is negative, it can reduce the noise below the vacuum noise level of 1, to a limit determined by the last term, representing thermal noise.
Indeed, it can be shown \cite{SI} that in this regime the noise level is bound from below by
\begin{equation}
	\bar S_{XX}^\mathrm{out}(\Omega)\gtrsim 1-\frac{\Gmeas}{\Gmeas+\Gdec},
	\label{eq:SqueezingBound}
\end{equation}
implying that large squeezing requires the measurement rate to significantly exceed the decoherence rate.
The photon collection inefficiencies discussed above reduce the squeezing further to
\begin{align}
	\bar S_{XX}^\mathrm{meas}(\Omega)&=\eta \bar S_{XX}^\mathrm{out}(\Omega)+(1-\eta) 1.
    \label{eq:MeasuredSpectrum}
\end{align}

\begin{figure}[tb]
\includegraphics[width=1\linewidth]{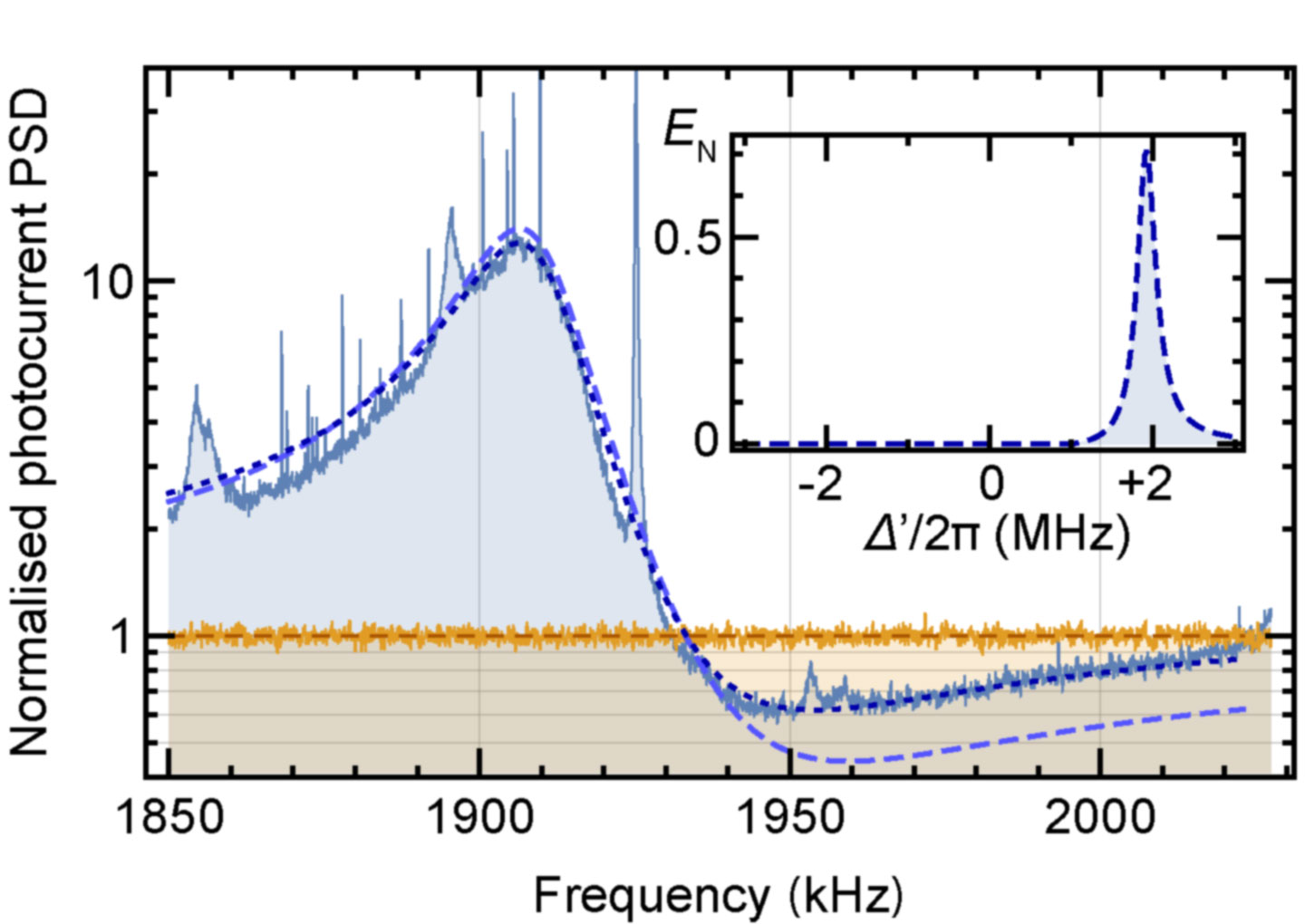}
\caption{Left: The strongest squeezing trace (the (3, 2) mode) with the theoretical model superimposed without any free parameter (dashed) and with adjusted cavity outcoupling and mirror noise (dotted).
Inset shows computed logarithmic negativity $E_\mathcal{N}$ of the ideally reachable entanglement between the mechanical mode and an optical mode extracted from the output beam by a filter centred at a frequency $\omega_\mathrm{f}=\oL-\Delta'$ off the laser carrier. 
}
\label{fig:squeezingandlogneg}
\end{figure}

We compare the data on the $(3,2)$ mode with the model \eqref{eq:MeasuredSpectrum}, supplied with independently determined parameters: $\kappa$, $\Delta$ and $g/2 \pi=580\unit{kHz}$ are extracted from OMIT traces, which yield a very high measurement rate of $\Gmeas/2\pi\approx 96\unit{kHz}$.
Mechanical frequency $\Om/2\pi=1928\unit{kHz}$ and damping $\Gm/2\pi=170\unit{mHz}$ are determined in cryogenic ringdown measurements.
The bath temperature $T=10\pm 0.4\unit{K}$ is obtained via comparison with a reference temperature using a frequency modulation calibration \cite{Gorodetsky2010}.
While the detection efficiency $\etaDet$ is determined by optical and photodetection losses, the cavity outcoupling efficiency hinges on the loss rate $\kT$ of the outcoupling mirror with respect to the total number of intracavity photons.
In a transfer matrix model \cite{Jayich2008, Wilson2009, SI} we calculate $\kT/2\pi =13{.}4\unit{MHz}$ from the known mirror  and membrane transmission and positions. 
Figure \ref{fig:squeezingandlogneg} shows a direct comparison of this zero-parameter model with the measured noise trace.
While the overall structure and signal-to-background level are well reproduced, the model predicts somewhat stronger squeezing.
We attribute this discrepancy to a combination of an overestimated collection efficiency $\eta$ (for example, due to membrane tilt), residual  frequency noise (most likely caused by substrate noise of the cavity mirrors) and contributions from neighbouring modes.
We find a better agreement if we allow an adjustment of the outcoupling efficiency and residual frequency noise, to better match the observed contrast and overall noise level, respectively.
Assuming $\kT/\kappa=80\%$ and a frequency noise level corresponding to a $25\%$ increase beyond shot noise in the absence of optomechanical coupling, we extract a cooling of the mechanical mode from an occupation $n\approx10^5$ to $n_\mathrm{eff}\approx 4{.}7$ ($4{.}3$ in the absence of mirror noise).

Interestingly, ponderomotive squeezing can be pictured to occur in two steps: a downconversion process first creates (or annihilates) an entangled pair of a red-sideband photon and a phonon. 
The latter is then converted to a blue-sideband photon in a swap process.
The resulting entanglement between red and blue sideband photons is measured as suppressed quantum fluctuations in a particular optical quadrature, at a particular sideband frequency.
This complementary perspective prompts us to theoretically evaluate the entanglement between the mechanical degree of freedom, and the light exiting the cavity \cite{Genes2008a,Palomaki2013a}.
By applying a spectral filter (e.~g.~a cavity) of width $\kappa'/2\pi=0{.}2\unit{MHz}$ to the output light, we define an isolated mode whose steady-state entanglement with the mechanical mode can be computed \cite{Genes2008a,SI}.
We calculate the logarithmic negativity $E_\mathcal{N}$ \cite{Eisert2001, Vidal2002,SI} for the parameters of Fig.~\ref{fig:squeezingandlogneg}, assuming that the reduced detection efficiency and effects of mirror noise can be avoided in an improved version of the experiment.
The resulting entanglement as the filter is tuned across the output light is shown in Fig.~\ref{fig:squeezingandlogneg}.
If the filtered mode coincides with the red (Stokes) sideband of the coupling laser we find a value as high as  $E_\mathcal{N}\approx0{.}7$.

In conclusion, we have demonstrated a robust, compact, multimode optomechanical system that exhibits strong optomechanical quantum correlations, as evidenced by significant ponderomotive squeezing.
Unprecedentedly large correlations are enabled by very low mechanical decoherence on one hand, and the highest yet realised detection efficiency in an optomechanics experiment on the one hand.
Crucially, a  phononic bandgap shield suppresses mechanical losses in a wide frequency range, so that quantum correlations can be observed with a large number of mechanical modes.

This system thus constitutes a promising platform for the realisation of a range of nonclassical mechanical states   \cite{Mancini2002, Hartmann2008, Bhattacharya2008b, Houhou2015}, as well as measurements of  displacements and forces beyond the standard quantum limit \cite{Woolley2013, Buchmann2015}.
The multimode nature of the optical and mechanical resonators, and the simplicity with which light, or the mechanics interface to other quantum systems---such as superconducting microwave circuits or atomic ensembles---multiplies the possible applications of this system as a multimode quantum interface \cite{Hammerer2009, Muschik2011, Barzanjeh2012, Tian2013, Andrews2013, Bagci2013}.

{\small $\hphantom{xxx}$\newline\noindent\textbf{Acknowledgements}
 This work was supported by the ERC grants Q-CEOM and INTERFACE, a starting grant from the Danish Council for Independent Research, the EU grants iQUOEMS and SIQS, the Defense Advanced Research Agency (DARPA) and the Carlsberg Foundation. The authors are grateful to K.~Usami and D.~J.~Wilson for input at an early stage of the project, and to E.~Belhage and A.~Barg for providing the dark field images.
 }

\bibliographystyle{naturemag_noURL}
\bibliography{references}

\begin{thebibliography}{10}
\expandafter\ifx\csname url\endcsname\relax
  \def\url#1{\texttt{#1}}\fi
\expandafter\ifx\csname urlprefix\endcsname\relax\def\urlprefix{URL }\fi
\providecommand{\bibinfo}[2]{#2}
\providecommand{\eprint}[2][]{\url{#2}}

\bibitem{Braginsky1992}
\bibinfo{author}{Braginsky, V.~B.} \& \bibinfo{author}{Khalili, F.~Y.}
\newblock \emph{\bibinfo{title}{Quantum Measurement}}
  (\bibinfo{publisher}{Cambridge University Press}, \bibinfo{year}{1992}).

\bibitem{Clerk2010}
\bibinfo{author}{Clerk, A.~A.}, \bibinfo{author}{Devoret, M.~H.},
  \bibinfo{author}{Girvin, S.~M.}, \bibinfo{author}{Marquardt, F.} \&
  \bibinfo{author}{Schoelkopf, R.~J.}
\newblock \bibinfo{title}{Introduction to quantum noise, measurement, and
  amplification}.
\newblock \emph{\bibinfo{journal}{Rev. Mod. Phys.}}
  \textbf{\bibinfo{volume}{82}}, \bibinfo{pages}{1155–1208}
  (\bibinfo{year}{2010}).

\bibitem{Aspelmeyer2014}
\bibinfo{author}{Aspelmeyer, M.}, \bibinfo{author}{Kippenberg, T.~J.} \&
  \bibinfo{author}{Marquardt, F.}
\newblock \bibinfo{title}{Cavity optomechanics}.
\newblock \emph{\bibinfo{journal}{Rev. Mod. Phys.}}
  \textbf{\bibinfo{volume}{86}}, \bibinfo{pages}{1391–1452}
  (\bibinfo{year}{2014}).

\bibitem{Murch2008}
\bibinfo{author}{Murch, K.~W.}, \bibinfo{author}{Moore, K.~L.},
  \bibinfo{author}{Gupta, S.} \& \bibinfo{author}{Stamper-Kurn, D.~M.}
\newblock \bibinfo{title}{Observation of quantum-measurement backaction with an
  ultracold atomic gas}.
\newblock \emph{\bibinfo{journal}{Nature Physics}}
  \textbf{\bibinfo{volume}{4}}, \bibinfo{pages}{561--564}
  (\bibinfo{year}{2008}).

\bibitem{Purdy2013}
\bibinfo{author}{Purdy, T.~P.}, \bibinfo{author}{Peterson, R.~W.} \&
  \bibinfo{author}{Regal, C.~A.}
\newblock \bibinfo{title}{Observation of radiation pressure shot noise on a
  macroscopic object}.
\newblock \emph{\bibinfo{journal}{Science}} \textbf{\bibinfo{volume}{339}},
  \bibinfo{pages}{801--804} (\bibinfo{year}{2013}).

\bibitem{Teufel2016}
\bibinfo{author}{Teufel, J.~D.}, \bibinfo{author}{Lecocq, F.} \&
  \bibinfo{author}{Simmonds, R.~W.}
\newblock \bibinfo{title}{Overwhelming thermomechanical motion with microwave
  radiation pressure shot noise}.
\newblock \emph{\bibinfo{journal}{Physical Review Letters}}
  \textbf{\bibinfo{volume}{116}}, \bibinfo{pages}{013602}
  (\bibinfo{year}{2016}).

\bibitem{Peterson2016}
\bibinfo{author}{Peterson, R.~W.} \emph{et~al.}
\newblock \bibinfo{title}{Laser cooling of a micromechanical membrane to the
  quantum backaction limit}.
\newblock \emph{\bibinfo{journal}{Physical Review Letters}}
  \textbf{\bibinfo{volume}{116}}, \bibinfo{pages}{063601}
  (\bibinfo{year}{2016}).

\bibitem{Vitali2007}
\bibinfo{author}{Vitali, D.} \emph{et~al.}
\newblock \bibinfo{title}{Optomechanical entanglement between a movable mirror
  and a cavity field}.
\newblock \emph{\bibinfo{journal}{Physical Review Letters}}
  \textbf{\bibinfo{volume}{98}}, \bibinfo{pages}{030405}
  (\bibinfo{year}{2007}).

\bibitem{Genes2008a}
\bibinfo{author}{Genes, C.}, \bibinfo{author}{Mari, A.},
  \bibinfo{author}{Tombesi, P.} \& \bibinfo{author}{Vitali, D.}
\newblock \bibinfo{title}{Robust entanglement of a micromechanical resonator
  with output optical fields}.
\newblock \emph{\bibinfo{journal}{Physical Review A}}
  \textbf{\bibinfo{volume}{78}} (\bibinfo{year}{2008}).

\bibitem{Khalili2012}
\bibinfo{author}{Khalili, F.~Y.}, \bibinfo{author}{Miao, H.},
  \bibinfo{author}{Safavi-Naeini, A.~H.}, \bibinfo{author}{Painter, O.} \&
  \bibinfo{author}{Chen, Y.}
\newblock \bibinfo{title}{Quantum back-action in measurements of zero-point
  mechanical oscillations}.
\newblock \emph{\bibinfo{journal}{Physical Review A}}
  \textbf{\bibinfo{volume}{86}}, \bibinfo{pages}{033840}
  (\bibinfo{year}{2012}).

\bibitem{Buchmann2016}
\bibinfo{author}{Buchmann, L.~F.}, \bibinfo{author}{Schreppler, S.},
  \bibinfo{author}{Kohler, J.}, \bibinfo{author}{Spethmann, N.} \&
  \bibinfo{author}{Stamper-Kurn, D.~M.}
\newblock \bibinfo{title}{Complex squeezing and force measurement beyond the
  standard quantum limit}.
\newblock \emph{\bibinfo{journal}{arXiv:1602.02141}}  (\bibinfo{year}{2016}).

\bibitem{Verlot2009}
\bibinfo{author}{Verlot, P.}, \bibinfo{author}{Tavernarakis, A.},
  \bibinfo{author}{Briant, T.}, \bibinfo{author}{Cohadon, P.-F.} \&
  \bibinfo{author}{Heidmann, A.}
\newblock \bibinfo{title}{Scheme to probe optomechanical correlations between
  two optical beams down to the quantum level}.
\newblock \emph{\bibinfo{journal}{Physical Review Letters}}
  \textbf{\bibinfo{volume}{102}}, \bibinfo{pages}{103601}
  (\bibinfo{year}{2009}).

\bibitem{Brooks2012}
\bibinfo{author}{Brooks, D. W.~C.} \emph{et~al.}
\newblock \bibinfo{title}{Non-classical light generated by quantum-noise-driven
  cavity optomechanics}.
\newblock \emph{\bibinfo{journal}{Nature}} \textbf{\bibinfo{volume}{488}},
  \bibinfo{pages}{476--480} (\bibinfo{year}{2012}).

\bibitem{Safavi-Naeini2012}
\bibinfo{author}{Safavi-Naeini, A.~H.} \emph{et~al.}
\newblock \bibinfo{title}{Observation of quantum motion of a nanomechanical
  resonator}.
\newblock \emph{\bibinfo{journal}{Phyical Review Letters}}
  \textbf{\bibinfo{volume}{108}}, \bibinfo{pages}{033602}
  (\bibinfo{year}{2012}).

\bibitem{Purdy2013a}
\bibinfo{author}{Purdy, T.~P.}, \bibinfo{author}{Yu, P.-L.},
  \bibinfo{author}{Peterson, R.~W.}, \bibinfo{author}{Kampel, N.~S.} \&
  \bibinfo{author}{Regal, C.~A.}
\newblock \bibinfo{title}{Strong optomechanical squeezing of light}.
\newblock \emph{\bibinfo{journal}{Physical Review X}}
  \textbf{\bibinfo{volume}{3}} (\bibinfo{year}{2013}).

\bibitem{Safavi-Naeini2013}
\bibinfo{author}{Safavi-Naeini, A.~H.} \emph{et~al.}
\newblock \bibinfo{title}{Squeezed light from a silicon micromechanical
  resonator}.
\newblock \emph{\bibinfo{journal}{Nature}} \textbf{\bibinfo{volume}{500}},
  \bibinfo{pages}{185} (\bibinfo{year}{2013}).

\bibitem{Weinstein2014}
\bibinfo{author}{Weinstein, A.~J.} \emph{et~al.}
\newblock \bibinfo{title}{Observation and interpretation of motional sideband
  asymmetry in a quantum electromechanical device}.
\newblock \emph{\bibinfo{journal}{Physical Review X}}
  \textbf{\bibinfo{volume}{4}}, \bibinfo{pages}{041003} (\bibinfo{year}{2014}).

\bibitem{Underwood2015}
\bibinfo{author}{Underwood, M.} \emph{et~al.}
\newblock \bibinfo{title}{Measurement of the motional sidebands of a
  nanogram-scale oscillator in the quantum regime}.
\newblock \emph{\bibinfo{journal}{Physical Review A}}
  \textbf{\bibinfo{volume}{92}}, \bibinfo{pages}{061801(R)}
  (\bibinfo{year}{2015}).

\bibitem{Sudhir2016}
\bibinfo{author}{Sudhir, V.} \emph{et~al.}
\newblock \bibinfo{title}{Appearance and disappearance of quantum correlations
  in measurement-based feedback control of a mechanical oscillator}.
\newblock \emph{\bibinfo{journal}{arXiv:1602.05942}}  (\bibinfo{year}{2016}).

\bibitem{Mancini2002}
\bibinfo{author}{Mancini, S.}, \bibinfo{author}{Giovannetti, V.},
  \bibinfo{author}{Vitali, D.} \& \bibinfo{author}{Tombesi, P.}
\newblock \bibinfo{title}{Entangling {M}acroscopic {O}scillators {E}xploiting
  {R}adiation {P}ressure}.
\newblock \emph{\bibinfo{journal}{Physical Review Letters}}
  \textbf{\bibinfo{volume}{88}}, \bibinfo{pages}{120401}
  (\bibinfo{year}{2002}).

\bibitem{Hartmann2008}
\bibinfo{author}{Hartmann, M.~J.} \& \bibinfo{author}{Plenio, M.~B.}
\newblock \bibinfo{title}{Steady state entanglement in the mechanical
  vibrations of two dielectric membranes}.
\newblock \emph{\bibinfo{journal}{Physical Review Letters}}
  \textbf{\bibinfo{volume}{101}}, \bibinfo{pages}{200503}
  (\bibinfo{year}{2008}).

\bibitem{Bhattacharya2008b}
\bibinfo{author}{Bhattacharya, M.} \& \bibinfo{author}{Meystre, P.}
\newblock \bibinfo{title}{Multiple membrane cavity optomechanics}.
\newblock \emph{\bibinfo{journal}{Physical Review A}}
  \textbf{\bibinfo{volume}{78}}, \bibinfo{pages}{041801}
  (\bibinfo{year}{2008}).

\bibitem{Houhou2015}
\bibinfo{author}{Houhou, O.}, \bibinfo{author}{Aissaoui, H.} \&
  \bibinfo{author}{Ferraro, A.}
\newblock \bibinfo{title}{Generation of cluster states in optomechanical
  quantum systems}.
\newblock \emph{\bibinfo{journal}{Physical Review A}}
  \textbf{\bibinfo{volume}{92}}, \bibinfo{pages}{063843}
  (\bibinfo{year}{2015}).

\bibitem{Woolley2013}
\bibinfo{author}{Woolley, M.~J.} \& \bibinfo{author}{Clerk, A.~A.}
\newblock \bibinfo{title}{Two-mode back-action-evading measurements in cavity
  optomechanics}.
\newblock \emph{\bibinfo{journal}{Physical Review A}}
  \textbf{\bibinfo{volume}{87}}, \bibinfo{pages}{063846}
  (\bibinfo{year}{2013}).

\bibitem{Buchmann2015}
\bibinfo{author}{Buchmann, L.~F.} \& \bibinfo{author}{Stamper-Kurn, D.~M.}
\newblock \bibinfo{title}{The quantum/classical transition in mediated
  interactions}.
\newblock \emph{\bibinfo{journal}{Annalen der Physik}}
  \textbf{\bibinfo{volume}{527}}, \bibinfo{pages}{156} (\bibinfo{year}{2015}).

\bibitem{Massel2012}
\bibinfo{author}{Massel, F.} \emph{et~al.}
\newblock \bibinfo{title}{Multimode circuit optomechanics near the quantum
  limit}.
\newblock \emph{\bibinfo{journal}{Nature Communications}}
  \textbf{\bibinfo{volume}{3}}, \bibinfo{pages}{987} (\bibinfo{year}{2012}).

\bibitem{Noguchi2016}
\bibinfo{author}{Noguchi, A.} \emph{et~al.}
\newblock \bibinfo{title}{Strong coupling in multimode quantum
  electromechanics}.
\newblock \emph{\bibinfo{journal}{arXiv:1602.01554}}  (\bibinfo{year}{2016}).

\bibitem{Spethmann2016}
\bibinfo{author}{Spethmann, N.}, \bibinfo{author}{Kohler, J.},
  \bibinfo{author}{Schreppler, S.}, \bibinfo{author}{Buchmann, L.} \&
  \bibinfo{author}{Stamper-Kurn, D.~M.}
\newblock \bibinfo{title}{Cavity-mediated coupling of mechanical oscillators
  limited by quantum back-action}.
\newblock \emph{\bibinfo{journal}{Nature Physics}}
  \textbf{\bibinfo{volume}{12}}, \bibinfo{pages}{27} (\bibinfo{year}{2016}).

\bibitem{Hammerer2010}
\bibinfo{author}{Hammerer, K.}, \bibinfo{author}{S{\o}rensen, A.~S.} \&
  \bibinfo{author}{Polzik, E.~S.}
\newblock \bibinfo{title}{Quantum interface between light and atomic
  ensembles}.
\newblock \emph{\bibinfo{journal}{Rev. Mod. Phys.}}
  \textbf{\bibinfo{volume}{82}}, \bibinfo{pages}{1041} (\bibinfo{year}{2010}).

\bibitem{Krauter2011}
\bibinfo{author}{Krauter, H.} \emph{et~al.}
\newblock \bibinfo{title}{Entanglement generated by dissipation and steady
  state entanglement of two macroscopic objects}.
\newblock \emph{\bibinfo{journal}{Physical Review Letters}}
  \textbf{\bibinfo{volume}{107}} (\bibinfo{year}{2011}).

\bibitem{Vasilakis2015}
\bibinfo{author}{Vasilakis, G.} \emph{et~al.}
\newblock \bibinfo{title}{Generation of a squeezed state of an oscillator by
  stroboscopic back-action-evading measurement}.
\newblock \emph{\bibinfo{journal}{Nature Physics}}
  \textbf{\bibinfo{volume}{11}}, \bibinfo{pages}{389￢ﾀﾓ392}
  (\bibinfo{year}{2015}).

\bibitem{Jayich2008}
\bibinfo{author}{Jayich, A.~M.} \emph{et~al.}
\newblock \bibinfo{title}{Dispersive optomechanics: a membrane inside a
  cavity}.
\newblock \emph{\bibinfo{journal}{New Journal of Physics}}
  \textbf{\bibinfo{volume}{10}}, \bibinfo{pages}{095008}
  (\bibinfo{year}{2008}).

\bibitem{Wilson2009}
\bibinfo{author}{Wilson, D.~J.}, \bibinfo{author}{Regal, C.~A.},
  \bibinfo{author}{Papp, S.~B.} \& \bibinfo{author}{Kimble, H.~J.}
\newblock \bibinfo{title}{Cavity optomechanics with stoichiometric sin films}.
\newblock \emph{\bibinfo{journal}{Physical Review Letters}}
  \textbf{\bibinfo{volume}{103}} (\bibinfo{year}{2009}).

\bibitem{SI}
\bibinfo{note}{See supplementary information.}

\bibitem{Tsaturyan2014}
\bibinfo{author}{Tsaturyan, Y.} \emph{et~al.}
\newblock \bibinfo{title}{{Demonstration of suppressed phonon tunneling losses
  in phononic bandgap shielded membrane resonators for high-$Q$
  optomechanics}}.
\newblock \emph{\bibinfo{journal}{Optics Express}}
  \textbf{\bibinfo{volume}{22}}, \bibinfo{pages}{6810} (\bibinfo{year}{2014}).

\bibitem{Wilson-Rae2011}
\bibinfo{author}{Wilson-Rae, I.} \emph{et~al.}
\newblock \bibinfo{title}{High-$q$ nanomechanics via destructive interference
  of elastic waves}.
\newblock \emph{\bibinfo{journal}{Physical Review Letters}}
  \textbf{\bibinfo{volume}{106}}, \bibinfo{pages}{047205}
  (\bibinfo{year}{2011}).

\bibitem{Villanueva2014}
\bibinfo{author}{Villanueva, L.~G.} \& \bibinfo{author}{Schmid, S.}
\newblock \bibinfo{title}{Evidence of surface loss as ubiquitous limiting
  damping mechanism in sin micro- and nanomechanical resonators}.
\newblock \emph{\bibinfo{journal}{Physical Review Letters}}
  \textbf{\bibinfo{volume}{113}} (\bibinfo{year}{2014}).

\bibitem{Weis2010}
\bibinfo{author}{Weis, S.} \emph{et~al.}
\newblock \bibinfo{title}{Optomechanically induced transparency}.
\newblock \emph{\bibinfo{journal}{Science}} \textbf{\bibinfo{volume}{330}},
  \bibinfo{pages}{1520--1523} (\bibinfo{year}{2010}).

\bibitem{Chakram2013}
\bibinfo{author}{Chakram, S.}, \bibinfo{author}{Patil, Y.~S.},
  \bibinfo{author}{Chang, L.} \& \bibinfo{author}{Vengalattore, M.}
\newblock \bibinfo{title}{Dissipation in ultrahigh quality factor {S}i{N}
  membrane resonators}.
\newblock \emph{\bibinfo{journal}{Physical Review Letters}}
  \textbf{\bibinfo{volume}{112}}, \bibinfo{pages}{127201}
  (\bibinfo{year}{2014}).

\bibitem{Palomaki2013a}
\bibinfo{author}{Palomaki, T.~A.}, \bibinfo{author}{Teufel, J.~D.},
  \bibinfo{author}{Simmonds, R.~W.} \& \bibinfo{author}{Lehnert, K.~W.}
\newblock \bibinfo{title}{Entangling mechanical motion with microwave fields}.
\newblock \emph{\bibinfo{journal}{Science}} \textbf{\bibinfo{volume}{342}},
  \bibinfo{pages}{710} (\bibinfo{year}{2013}).

\bibitem{Verhagen2012}
\bibinfo{author}{Verhagen, E.}, \bibinfo{author}{Deléglise, S.},
  \bibinfo{author}{Weis, S.}, \bibinfo{author}{Schliesser, A.} \&
  \bibinfo{author}{Kippenberg, T.~J.}
\newblock \bibinfo{title}{Quantum-coherent coupling of a mechanical oscillator
  to an optical cavity mode}.
\newblock \emph{\bibinfo{journal}{Nature}} \textbf{\bibinfo{volume}{482}},
  \bibinfo{pages}{63–67} (\bibinfo{year}{2012}).

\bibitem{Wilson2015}
\bibinfo{author}{Wilson, D.~J.} \emph{et~al.}
\newblock \bibinfo{title}{Measurement-based control of a mechanical oscillator
  at its thermal decoherence rate}.
\newblock \emph{\bibinfo{journal}{Nature}} \textbf{\bibinfo{volume}{524}},
  \bibinfo{pages}{325–329} (\bibinfo{year}{2015}).

\bibitem{Fabre1994}
\bibinfo{author}{Fabre, C.} \emph{et~al.}
\newblock \bibinfo{title}{Quantum-noise reduction using a cavity with a movable
  mirror}.
\newblock \emph{\bibinfo{journal}{Phys. Rev. A}}  (\bibinfo{year}{1994}).

\bibitem{Mancini1994}
\bibinfo{author}{Mancini, S.} \& \bibinfo{author}{Tombesi, P.}
\newblock \bibinfo{title}{Quantum noise reduction by radiation pressure}.
\newblock \emph{\bibinfo{journal}{Phys. Rev. A}} \bibinfo{pages}{4055}
  (\bibinfo{year}{1994}).

\bibitem{Schliesser2006}
\bibinfo{author}{Schliesser, A.}, \bibinfo{author}{Del'Haye, P.},
  \bibinfo{author}{Nooshi, N.}, \bibinfo{author}{Vahala, K.} \&
  \bibinfo{author}{Kippenberg, T.}
\newblock \bibinfo{title}{Radiation pressure cooling of a micromechanical
  oscillator using dynamical backaction}.
\newblock \emph{\bibinfo{journal}{Physical Review Letters}}
  \textbf{\bibinfo{volume}{97}}, \bibinfo{pages}{243905}
  (\bibinfo{year}{2006}).

\bibitem{Arcizet2006a}
\bibinfo{author}{Arcizet, O.}, \bibinfo{author}{Cohadon, P.-F.},
  \bibinfo{author}{Briant, T.}, \bibinfo{author}{Pinard, M.} \&
  \bibinfo{author}{Heidmann, A.}
\newblock \bibinfo{title}{Radiation-pressure cooling and optomechanical
  instability of a micromirror}.
\newblock \emph{\bibinfo{journal}{Nature}} \textbf{\bibinfo{volume}{444}},
  \bibinfo{pages}{71–74} (\bibinfo{year}{2006}).

\bibitem{Gorodetsky2010}
\bibinfo{author}{Gorodetsky, M.~L.}, \bibinfo{author}{Schliesser, A.},
  \bibinfo{author}{Anetsberger, G.}, \bibinfo{author}{Deleglise, S.} \&
  \bibinfo{author}{Kippenberg, T.~J.}
\newblock \bibinfo{title}{{D}etermination of the vacuum optomechanical coupling
  rate using frequency noise calibration}.
\newblock \emph{\bibinfo{journal}{Optics Express}}
  \textbf{\bibinfo{volume}{18}}, \bibinfo{pages}{23236} (\bibinfo{year}{2010}).

\bibitem{Eisert2001}
\bibinfo{author}{Eisert, J.} \& \bibinfo{author}{Briegel, H.~J.}
\newblock \bibinfo{title}{Schmidt measure as a tool for quantifying
  multiparticle entanglement}.
\newblock \emph{\bibinfo{journal}{Physical Review A}}
  \textbf{\bibinfo{volume}{64}} (\bibinfo{year}{2001}).

\bibitem{Vidal2002}
\bibinfo{author}{Vidal, G.} \& \bibinfo{author}{Werner, R.~F.}
\newblock \bibinfo{title}{Computable measure of entanglement}.
\newblock \emph{\bibinfo{journal}{Physical Review A}}
  \textbf{\bibinfo{volume}{65}} (\bibinfo{year}{2002}).

\bibitem{Hammerer2009}
\bibinfo{author}{Hammerer, K.}, \bibinfo{author}{Aspelmeyer, M.},
  \bibinfo{author}{Polzik, E.~S.} \& \bibinfo{author}{Zoller, P.}
\newblock \bibinfo{title}{{E}stablishing {E}instein-{P}odolsky-{R}osen
  {C}hannels between {N}anomechanics and {A}tomic {E}nsembles}.
\newblock \emph{\bibinfo{journal}{Phys. Rev. Lett.}}
  \textbf{\bibinfo{volume}{102}}, \bibinfo{pages}{020501}
  (\bibinfo{year}{2009}).

\bibitem{Muschik2011}
\bibinfo{author}{Muschik, C.~A.}, \bibinfo{author}{Krauter, H.},
  \bibinfo{author}{Hammerer, K.} \& \bibinfo{author}{Polzik, E.~S.}
\newblock \bibinfo{title}{Quantum information at the interface of light with
  atomic ensembles and micromechanical oscillators}.
\newblock \emph{\bibinfo{journal}{Quantum Information Processing}}
  \textbf{\bibinfo{volume}{10}}, \bibinfo{pages}{839￢ﾀﾓ863}
  (\bibinfo{year}{2011}).

\bibitem{Barzanjeh2012}
\bibinfo{author}{Barzanjeh, S.}, \bibinfo{author}{Abdi, M.},
  \bibinfo{author}{Milburn, G.~J.}, \bibinfo{author}{Tombesi, P.} \&
  \bibinfo{author}{Vitali, D.}
\newblock \bibinfo{title}{Reversible optical-to-microwave quantum interface}.
\newblock \emph{\bibinfo{journal}{Physical Review Letters}}
  \textbf{\bibinfo{volume}{109}}, \bibinfo{pages}{130503}
  (\bibinfo{year}{2012}).

\bibitem{Tian2013}
\bibinfo{author}{Tian, L.}
\newblock \bibinfo{title}{Robust photon entanglement via quantum interference
  in optomechanical interfaces}.
\newblock \emph{\bibinfo{journal}{Phyical Review Letters}}
  \textbf{\bibinfo{volume}{110}}, \bibinfo{pages}{233602}
  (\bibinfo{year}{2013}).

\bibitem{Andrews2013}
\bibinfo{author}{Andrews, R.~W.} \emph{et~al.}
\newblock \bibinfo{title}{Reversible and efficient conversion between microwave
  and optical light}.
\newblock \emph{\bibinfo{journal}{Nature Physics}}
  \textbf{\bibinfo{volume}{10}}, \bibinfo{pages}{321} (\bibinfo{year}{2014}).

\bibitem{Bagci2013}
\bibinfo{author}{Bagci, T.} \emph{et~al.}
\newblock \bibinfo{title}{Optical detection of radio waves through a
  nanomechanical transducer}.
\newblock \emph{\bibinfo{journal}{Nature}} \textbf{\bibinfo{volume}{507}},
  \bibinfo{pages}{81} (\bibinfo{year}{2014}).

\end{thebibliography}

\onecolumngrid

\setcounter{figure}{0}%
\setcounter{equation}{0}%
\setcounter{section}{0}
\renewcommand \theequation {A\arabic{equation}}%
\renewcommand \thefigure {A\arabic{figure}}
\renewcommand \thesection {A\arabic{section}}

\clearpage
\newpage

\begin{center}
\vspace{0.9in}
\Large{
\textbf{Supporting information}}
\end{center}
\vspace{0in}

\setcounter{tocdepth}{3}
\tableofcontents

\section{Theory}

\subsection{The Transfer Matrix Model}
\label{sec:transf-matr-model}

In order to describe a membrane-in-the-middle system with the canonical optomechanical theory, one must deploy a mapping from the actual (dispersive) geometry where the mechanically compliant part resides between the two cavity mirrors to the ``moving end mirror''-geometry of the canonical optomechanical system. Such a mapping is offered by the transfer matrix model, introduced in \cite{Jayich2008}. 

\begin{figure}[h]
  \centering \includegraphics[width=0.5\textwidth]{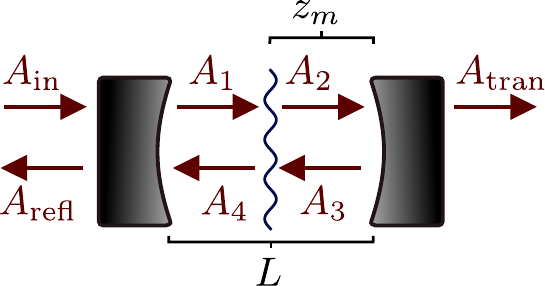}
  \caption{Overview of the fields used in the calculation.}
  \label{fig:TMMoverview}
\end{figure}

Defining four intra-cavity fields as shown in Figure \ref{fig:TMMoverview} and denoting the incoupler mirror as mirror 1, with associated amplitude transmission and reflection coefficients $t_1$ and $r_1$ (and similarly for mirror 2 and the membrane), we may describe the system as a whole with the following matrix equation:
\begin{equation}
  \label{eq:TMM_matrix}
  \begin{pmatrix}
    A_1\\
    A_2\\
    A_3\\
    A_4\\
  \end{pmatrix}
  =
  \begin{pmatrix}
    -1 & r_1\e^{\mathrm{i}k(L-z_m)} & 0 & 0 \\
    r_m\e^{\mathrm{i}k(L-z_m)} & -1 & 0 & \mathrm{i}t_m\e^{\mathrm{i}kz_m}\\
    \mathrm{i}t_m\e^{\mathrm{i}k(L-z_m)} & 0 & -1 & r_m\e^{\mathrm{i}kz_m} \\
    0 & 0 & r_2\e^{\mathrm{i}kz_m} & -1 \\
  \end{pmatrix}
  \begin{pmatrix}
    -A_\text{in}\\
    0\\
    0\\
    0\\
  \end{pmatrix},
\end{equation}
where $k$ is the wavenumber of the incoming laser light. We take $A_\text{in}=1$. For a given $k$, all intra-cavity fields are thus determined. The reflected and transmitted fields are given by
\begin{align}
    A_\text{tran} &= \mathrm{i}t_2A_3\e^{\mathrm{i}kz_m},\\
  A_\text{refl} &= \mathrm{i}t_1A_2\e^{\mathrm{i}k(L-z_m)}+r_1A_\text{in}.
\end{align}
The cavity resonance $k$-values may be found from the condition $\text{Im}(A_2)=0$.

In order to compute the solutions to the Heisenberg-Langevin equations, we need to determine $\kappa$, $g_0$, and $\eta_c$. For the latter two, the following analytic expressions may be used:
\begin{align}
  \eta_c &= \frac{|t_2|^2|A_3|^2}{|t_2|^2|A_3|^2+|t_1|^2|A_2|^2},\\
  g_0 &= x_\text{zpf}\omega_L\frac{|A_1|^2+|A_2|^2-|A_3|^2-|A_4|^2}{(L-z_m)(|A_1|^2+|A_2|^2)+z_m(|A_3|^2+|A_4|^2)},\\  
\end{align}
where $\omega_L = ck$.
For $\kappa$, we have not found an analytic expression. Instead, one may vary $k$ around the resonance and fit a Lorentzian to $A_\text{trans}$, thereby extracting the cavity linewidth. 

\begin{figure}[h]
  \centering
  \includegraphics[width=0.95\textwidth]{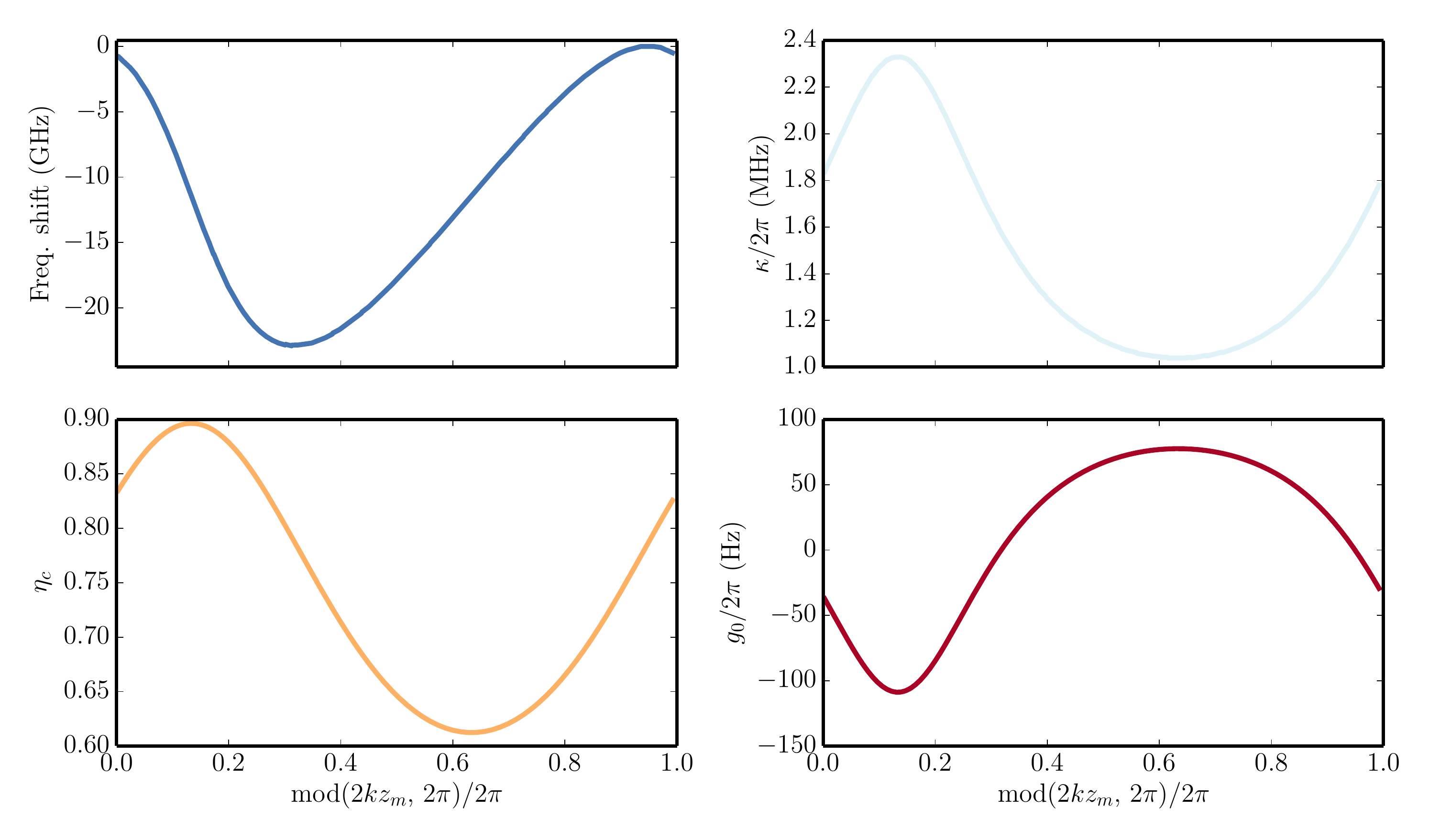}
  \caption{Model predictions for parameters equal to the ones measured and used in the experiment. The $g_0$ is for a mode with $\Omega_m=2\pi\times1.92$ MHz and (physical) mass $m=62$ ng, i.e. similar to the (3, 2)-mode of the experiment.}
  \label{fig:COWplot}
\end{figure}

In the dispersive geometry, $\kappa$, $g_0$, and $\eta_c$ depend periodically on $k$, which may be understood from the fact that the phase shift imparted on the light by the membrane depends on the relative position of the membrane in the intra-cavity standing wave. In particular, the period $T$ fulfils that  $T/2\pi=2kz_m$. In Figure \ref{fig:COWplot} we plot the periodic variation of $\kappa$, $\eta_c$, and $g_0$ with $2kz_m$. Also shown is the cavity resonance frequency shift from the empty cavity value. The frequency shift offers an easily measurable way of determining the $2kz_m$ value of a given resonance, a quantity otherwise difficult to measure.

In the experiment, the lack of a tunable membrane position necessitates a method to ascertain the relative placement of the membrane to the intracavity standing light wave, since many vital experimental parameters such as optomechanical coupling and cavity linewidth depend on this \cite{Jayich2008}. As the presence of the membrane causes large cavity resonance frequency shifts, $\Delta f_\text{cav}$, with a period of $2kz_m$  from the normal linear behaviour of en empty cavity, we rely on this shift to easily map out the relative membrane-standing wave position.  In Fig. 1(c) these shifts are displayed together with the accompanying periodic modulation of cavity linewidth. 

\subsection{Langevin equations}

The dynamics are described by the linearised equations of motion for the quadratures $X(t), Y(t)$ of the intracavity light field, and the mechanical motion $q(t), p(t)$
\begin{align}
        \label{eq:Appendix-HLE-X}
	\dot X(t) &=-\frac{\kappa}{2} X(t)-\Delta Y(t)+\sqrt{\kappa} X_\mathrm{in}(t)\\ \label{eq:Appendix-HLE-Y}
	\dot Y(t) &=-\frac{\kappa}{2} Y(t)+\Delta X(t)+2 g q(t)+\sqrt{\kappa} Y_\mathrm{in}(t)\\ \label{eq:Appendix-HLE-q}
	\dot q(t)&=\Om p(t)\\ \label{eq:Appendix-HLE-p}
	\dot p(t)&=-\Om q(t)-\Gm p(t) +2g X(t)+\sqrt{\Gm} p_\mathrm{in}(t) 
\intertext{The cavity output, in terms of quadratures can be calculated using the input-output relations } \label{eq:Appendix-IO-X}
	X_\mathrm{out}(t)&=\sqrt{\etaDet}\left(X_\mathrm{in}(t)-\sqrt{\kappa} X(t) \right)+\sqrt{1-\etaDet}X_\mathrm{vac}(t)\\ \label{eq:Appendix-IO-Y}
	Y_\mathrm{out}(t)&=\sqrt{\etaDet}\left(Y_\mathrm{in}(t)-\sqrt{\kappa} Y(t) \right)+\sqrt{1-\etaDet}Y_\mathrm{vac}(t),
\end{align}
where we have taken into account that losses (inside or outside the cavity) can occur, leading to a detection efficiency $\etaDet<1$.

\subsection{Squeezing spectra}

\subsubsection{General solution}

The squeezing spectra may be calculated from the four equations of motions \eqref{eq:Appendix-HLE-X}-\eqref{eq:Appendix-HLE-p} combined with the input-output relations \eqref{eq:Appendix-IO-X} and \eqref{eq:Appendix-IO-Y}. Cast in matrix form in the frequency domain,
\begin{align}
  \label{eq:Appendix-matrixeomXYqp}
  w(\Omega) &= (-\mathrm{i}\Omega\mathbb{1}+M')^{-1}w^\text{in}(\Omega),\\
            &=: L(\Omega) w^\text{in}(\Omega),
\end{align}
where
\begin{equation}
  \label{eq:Appendix-defofwandw_in}
  w(\Omega) =
  \begin{pmatrix}
    X(\Omega) \\
    Y(\Omega) \\
    q(\Omega) \\
    p(\Omega)
  \end{pmatrix}, \quad
  w_\text{in}(\Omega) =
  \begin{pmatrix}
    \sqrt{\kappa}\ X_\text{in}(\Omega) \\
    \sqrt{\kappa}\   Y_\text{in}(\Omega) \\
    0 \\
    \sqrt{\Gm}\ p_\text{in}(\Omega)    
  \end{pmatrix}
\end{equation}
and
\begin{equation}
 M'=\begin{pmatrix}
  		\kappa/2				& \Delta					&0 		&	0 \\
  		-\Delta					& \kappa/2					& 2 \ecr 				&	0 \\
		0						& 0							& 0					& -\Om \\
	2\ecr		&0							& \Om				& \Gm
\end{pmatrix}.
\end{equation}
Defining an input noise covariance matrix, $W(\Omega, \Omega')$, by
\begin{equation}
  \label{eq:1}
  W_{ij}(\Omega, \Omega') := \mean{w^\text{in}_i(\Omega) w^\text{in}_j(\Omega')},
\end{equation}
the power spectral density $S$, given for a Hermitian operator $A$ by
\begin{equation}
  \label{eq:2}
  S_{AA}(\Omega) = \int \mathrm{d}\Omega' \mean{A(\Omega)A(\Omega')},
\end{equation}
can then be computed for our operators of interest as
\begin{equation}
  \label{eq:3}
  S_{w_iw_j}(\Omega) = \int \mathrm{d}\Omega' \left(L^*(\Omega)W(\Omega, \Omega')L^{T}(\Omega') \right)_{ij}.
\end{equation}
Equation \eqref{eq:3} leads to lengthy expressions that simplify significantly upon symmetrisation. When considering the symmetrised spectrum, $\bar{S}_{w_iw_i}$, given by
\begin{equation}
  \label{eq:Appendix-symmetrised_S}
  \bar{S}_{w_iw_i}(\Omega)=\frac{1}{2}[S_{w_iw_i}(\Omega)+S_{w_iw_i}(-\Omega)]
\end{equation}
and making the Markov approximation for the thermal bath, we have that
\begin{equation}
  \label{eq:Appendix-simpelsymm}
  \bar{S}_{w_iw_i}(\Omega) = \sum_{j=1}^4 L_{ij}(\Omega)L_{ij}(-\Omega)D_{jj},
\end{equation}
where the symmetrised covariance matrix $D$ is given by
\begin{equation}
 \label{eq:Appendix-noisematrix}
  D = 
  \begin{pmatrix}
    \kappa/2 & 0 & 0 & 0 \\
    0 & \kappa/2 & 0 & 0 \\
    0 & 0 & 0  & 0 \\
     0 & 0 &  0 & 2\Gm\nth\\  
  \end{pmatrix},
\end{equation}
where we have made the approximation that $\nth+1/2\approx \nth$. From here, it is easy to extend to the spectra of output fluctuations, which is essentially added noise to the intra-cavity spectral densities. Of main interest to us is
\begin{equation}
  \label{eq:Appendix-outputspectrum} 
  \bar{S}^\text{out}_{XX}(\Omega) = \etaDet\kappa \bar{S}_{XX}(\Omega) + 1- \etaDet\kappa[L_{11}(\Omega)+L_{11}(-\Omega)].
\end{equation}
Note that with our convention for the input noise operators, $1$ corresponds to the shot noise level.

\subsubsection{Approximate solution}

More intuitive expressions can be derived by writing \eqref{eq:Appendix-HLE-X} as
\begin{equation}
 	X(\Omega)=\frac{4  g u }{\kappa} q(\Omega)+\frac{2}{\sqrt{\kappa}}\left(u Y_\mathrm{in}(\Omega)+v X_\mathrm{in}(\Omega) \right)
	\label{eq:AppendixApprox01}
\end{equation}
using the abbreviations 
\begin{align}
	u&\equiv\frac{-2 \Delta }{4 \Delta^2 +(\kappa-2 i\Omega)^2}\kappa\\
	v&\equiv\frac{\kappa-2i\Omega}{4 \Delta^2 +(\kappa-2 i\Omega)^2} \kappa,
\end{align}
 and the mechanical equations of motion \eqref{eq:Appendix-HLE-q},\eqref{eq:Appendix-HLE-p} as 
\begin{equation}
	\chim(\Omega)^{-1} q(\Omega)=2g X(\Omega)+\sqrt{\Gm} p_\mathrm{in}(\Omega)
	\label{eq:AppendixApprox02}
\end{equation}
 with a mechanical susceptibility
\begin{equation}
  \chim(\Omega)=\frac{\Om}{\Om^2-\Omega^2-i \Gm \Omega} .
\end{equation}
Substituting \eqref{eq:AppendixApprox01} into \eqref{eq:AppendixApprox02} yields
\begin{equation}
	\chieff(\Omega)^{-1} q(\Omega)=4\frac{g}{ \sqrt{\kappa}}\left(u Y_\mathrm{in}(\Omega)+v X_\mathrm{in}(\Omega)\right)+\sqrt{\Gm} p_\mathrm{in}(\Omega),
\end{equation}
where we have introduced the usual  \cite{Aspelmeyer2014} effective mechanical susceptibility $\chieff(\Omega)$.
Resubstitution into \eqref{eq:AppendixApprox01} yields
\begin{equation}
	X(\Omega)=\left(16 \frac{g^2 u}{\kappa}\chieff(\Omega) +2\right)\frac{1}{\sqrt{\kappa}}\left(u Y_\mathrm{in}(\Omega)+v X_\mathrm{in}(\Omega)\right)+\frac{4 gu}{\kappa}\chieff(\Omega)\sqrt{\Gm} p_\mathrm{in}(\Omega)
\end{equation}
and 
\begin{align}
	X_\mathrm{out}(\Omega)&=-\left(16 \frac{g^2 u}{\kappa}\chieff(\Omega) +2\right)\left(u Y_\mathrm{in}(\Omega)+v X_\mathrm{in}(\Omega)\right)+X_\mathrm{in}-\frac{4 gu}{\sqrt{\kappa}}\chieff(\Omega)\sqrt{\Gm} p_\mathrm{in}(\Omega)\nonumber\\
	&=-\left(16 \frac{g^2 u^2}{\kappa}\chieff(\Omega) +2 u   \right)Y_\mathrm{in}(\Omega)
		-\left(16 \frac{g^2 u v}{\kappa}\chieff(\Omega) +2 v-1\right)X_\mathrm{in}(\Omega)
		+mech.
\end{align}
The symmetrised power spectral density of the output amplitude quadrature fluctuations is then given by 
\begin{align}
	\bar S_{XX}^\mathrm{out}(\Omega)&=
		1
		+\frac{32 g^2}{\kappa}\mathrm{Re}\left\{\chieff(\Og) u(2u^2+2v^2-v)\right\}
		+\left(\frac{16 g^2}{\kappa}\right)^2 \left|\chieff(\Og)\right|^2 |u(u+v)|^2+\nonumber\\
		&{}\qquad\qquad		+\frac{16 g^2}{\kappa}\left|\chieff(\Og)\right|^2 |u|^2 4\Gm n,
\end{align}
where the four terms are, respectively, imprecision noise, correlation term, quantum backaction and thermal noise.
Squeezing can occur of the correlation term is negative.

For a fast optical cavity with $\kappa \gg |\Delta|, \Og$ and a high cooperativty $4 g^2/\kappa\Gm\gg1$, these expressions are well approximated by
\begin{align}
	\bar S_{XX}^\mathrm{out}(\Omega)&\approx
		1-2\frac{8 \Delta}{\kappa}\frac{4 g^2}{\kappa}\mathrm{Re}\left\{\chieff(\Og) \right\}
		+\left(\frac{8 \Delta}{\kappa}\right)^2 \left(\frac{4 g^2}{\kappa}\right)^2 \left|\chieff(\Og)\right|^2
		+\left(\frac{2 \Delta}{\kappa}\right)^2 \left(\frac{16 g^2}{\kappa}\right) \left|\chieff(\Og)\right|^2 4\Gm n \nonumber\\
		&\approx
		1-2\frac{8 \Delta}{\kappa}\frac{4 g^2}{\kappa}\mathrm{Re}\left\{\chieff(\Og) \right\}
		+\left(\frac{8 \Delta}{\kappa}\right)^2 \frac{4 g^2}{\kappa} \left|\chieff(\Og)\right|^2
		\left(\frac{4 g^2}{\kappa} +n \Gm \right).
	\label{eq:AppendixApproxFinal}
\end{align}
Introducing the abbreviations $\Gmeas={4 g^2}/{\kappa}$ and $\theta= {8\Delta}/{\kappa}$ this may be rewritten as
\begin{align}
	\bar S_{XX}^\mathrm{out}(\Omega)&\approx
		1-2\theta \Gmeas\mathrm{Re}\left\{\chieff(\Og) \right\}
		+\theta^2 \Gmeas \left|\chieff(\Og)\right|^2
		\left(\Gmeas +n \Gm \right).
	\label{eq:AppendixApproxFinal}
\end{align}
\subsubsection{Squeezing bound}

Intuition suggests a lower bound for the attainable squeezing,
\begin{equation}
	\bar S_{XX}^\mathrm{out}(\Omega)\gtrsim 1-\frac{\Gmeas}{\Gmeas+\Gdec},
	\label{eq:AppendixSqueezingBound}
\end{equation}
which holds within the above approximation,
\begin{align}
	\bar S_{XX}^\mathrm{out}(\Omega)\approx
		1-2\theta\Gmeas\mathrm{Re}\left\{\chieff(\Og) \right\}
		+\theta^2 \Gmeas \left|\chieff(\Og)\right|^2
		\left(\Gmeas+\Gdec \right)
		&\geq 1-\frac{\Gmeas}{\Gmeas+\Gdec}
		 \nonumber\\
		\frac{\Gmeas}{\Gmeas+\Gdec} -2\theta\Gmeas\mathrm{Re}\left\{\chieff(\Og) \right\}
		+\theta^2 \Gmeas \left|\chieff(\Og)\right|^2
		\left(\Gmeas+\Gdec \right)
		&\geq0
		 \nonumber\\
		\frac{1}{\left|\chieff(\Og)\right|^2} -2\theta \frac{\mathrm{Re}\left\{\chieff(\Og) \right\}}{\left|\chieff(\Og)\right|^2}\left(\Gmeas+\Gdec \right)
		+\theta^2  
		\left(\Gmeas+\Gdec \right)^2
		&\geq0	
		\nonumber\\
		\frac{(\Omega_\mathrm{eff}^2-\Og^2)^2+\Geff^2\Og^2}{\Om^2} -2\theta \frac{\Omega_\mathrm{eff}^2-\Og^2}{\Om}\left(\Gmeas+\Gdec \right)
		+\theta^2  
		\left(\Gmeas+\Gdec \right)^2
		&\geq0
		\nonumber\\
		\Geff^2\Og^2+\left( (\Omega_\mathrm{eff}^2-\Og^2)-\theta\Om\left(\Gmeas+\Gdec \right)  \right)^2
		&\geq0.
	%
\end{align}

\begin{figure*}[htb]
\includegraphics[width=0.65\linewidth]{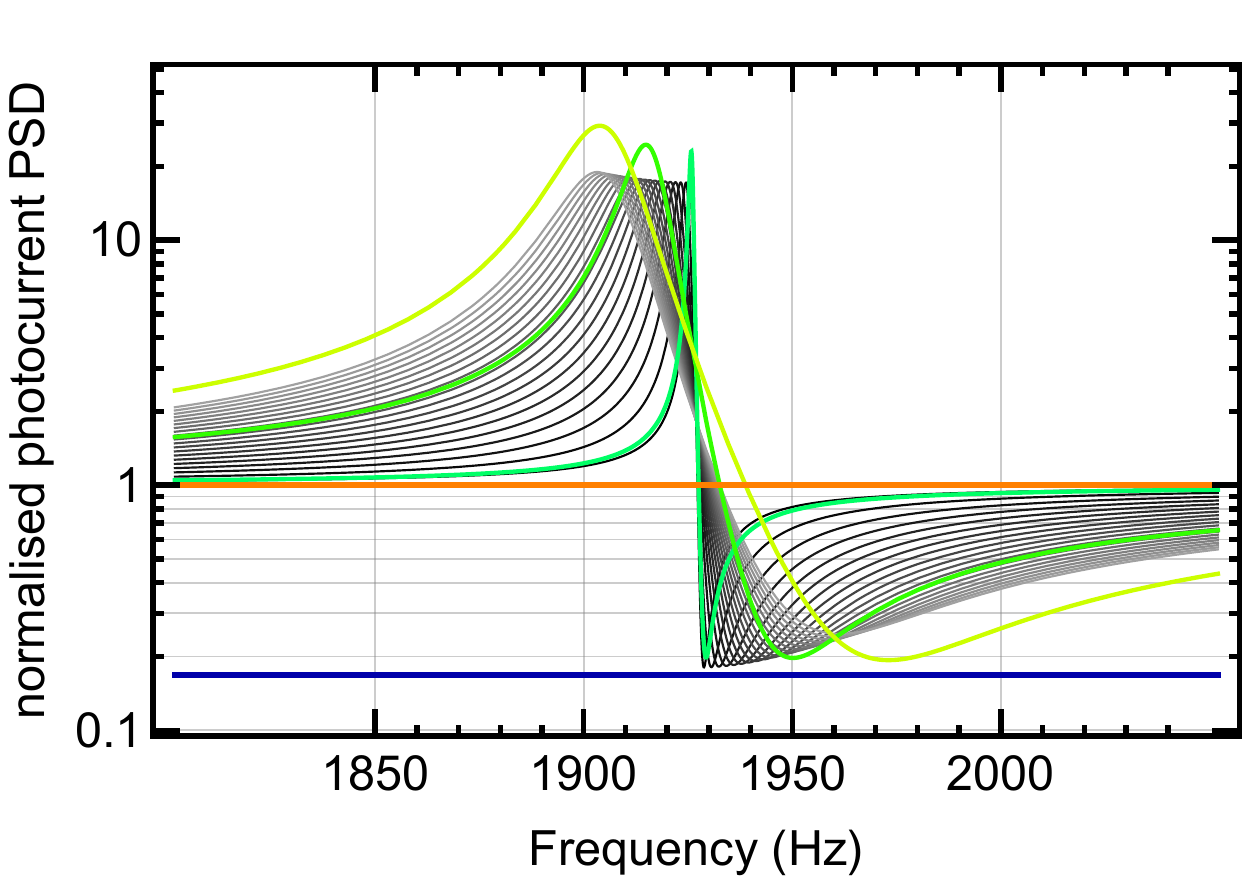}
\caption{Full model's prediction of photocurrent power spectral density, normalised to shot noise, with parameters as in figure \ref{fig:squeezingandlogneg} of the main text, but with maximum overcoupling of the cavity $\kT/\kappa\rightarrow 1$ and detection efficiency $\etaDet\rightarrow 1$, for  different detunings $\Delta/2\pi\in\{-2{.}0, -1{.}9, -1{.}8, \ldots, -0{.}1\}$ \text{MHz} (bright to dark grey). 
For comparison, predictions of the simplified model \eqref{eq:AppendixApproxFinal} are also shown for selected detunings ($\Delta/2\pi\in\{-2{.}0, -1{.}0, -0{.}1\}$ \text{MHz}). Orange line is vacuum noise and blue line the lower bound $1-\Gmeas/(\Gmeas+\Gdec)$.}
\label{f:fullModel}
\end{figure*}

This lower bound can be closely approached when a sufficiently small detuning is chosen.
Figure \ref{f:fullModel} shows a comparison of the full model \eqref{eq:Appendix-outputspectrum} for different detunings in comparison to approximation \eqref{eq:AppendixApproxFinal} and the limit of 	\eqref{eq:AppendixSqueezingBound}. 
The agreement between the full and the simplified model is particularly good for  small detunings ($\kappa/\Delta\gtrsim10$) as expected.

\subsection{Entanglement}

From the cavity output, one mode is isolated using a filter--essentially another optical cavity--of spectral width $\kappa'$, and detuned from the laser by the frequency $\Delta'$. This isolated mode evolves as
\begin{align}
	\dot x(t) &=-\frac{\kappa'}{2} x(t)-\Delta' y(t)+\sqrt{\kappa'} X_\mathrm{out}(t)\\
	\dot y(t) &=-\frac{\kappa'}{2} y(t)+\Delta' x(t)+\sqrt{\kappa'} Y_\mathrm{out}(t).
\end{align}
The linearised dynamics of the full system can be written as
\begin{align}
\frac{d}{dt}
    \begin{pmatrix} X \\  Y \\  q \\  p  \\ x \\ y \end{pmatrix} &=
    \begin{pmatrix}
  		-\kappa/2				& -\Delta					&0 					&	0		&	0		&	0 \\
  		+\Delta				& -\kappa/2				& 2\ecr 				&	0 		&	0		&	0\\
		0						& 0						& 0					& +\Om 	&	0		&	0\\
		2\ecr					&0							& -\Om				& -\Gm 	&	0		&	0\\
		-\sqrt{\kappa'\eta\kappa}	&0					& 0					& 0		&	-\kappa'/2		&	-\Delta'\\
		0						&-\sqrt{\kappa'\eta\kappa}	& 0				& 0		&	+\Delta'		&	-\kappa'/2
\end{pmatrix}
    \begin{pmatrix} X \\  Y \\  q \\  p  \\ x \\ y \end{pmatrix} +
    \begin{pmatrix} \sqrt{\kappa}X_\mathrm{in} \\  \sqrt{\kappa}Y_\mathrm{in} \\  0 \\  \sqrt{\Gm}p_\mathrm{in}  \\ \sqrt{\eta \kappa'}X_\mathrm{in} +\sqrt{(1-\eta) \kappa'}X_\mathrm{vac}  \\ \sqrt{\eta \kappa'}Y_\mathrm{in} +\sqrt{(1-\eta) \kappa'}Y_\mathrm{vac} \end{pmatrix},\nonumber
\end{align}
which has the form
\begin{equation}
  \frac{d}{dt} v(t)=A v(t)+v_\mathrm{in}(t).
\end{equation}
The input fluctuations $v_\mathrm{in}(t)$ are fully described by a covariances matrix $D$ given by
\begin{equation}
2D=    \begin{pmatrix}
  		\kappa				& 0						&0 					&	0		&	\sqrt{\kappa'\eta\kappa}		&	0 \\
  		0						& \kappa					& 						&	0 		&	0		&	\sqrt{\kappa'\eta\kappa}\\
		0						& 0						& 0					&  	0		&	0		&	0\\
		0						&0							& 						& 4 n \Gm 	&	0		&	0\\
		\sqrt{\kappa'\eta\kappa}	&0					& 0					& 0		&	\kappa'	&	0\\
		0						&\sqrt{\kappa'\eta\kappa}	& 0				& 0		&	0		&	\kappa'
\end{pmatrix}
\end{equation}
which can be computed from the known correlation functions (we have taken the classical limit for the mechanical bath). Then, for a stable system, the steady-state covariance matrix $V$ of $v(t)$ can be computed by solving the Lyapunov equation
\begin{equation}
  A V + V A^{T}=-D.
\end{equation}
The resulting correlation matrix can be written in block form
\begin{equation}
  V = \begin{pmatrix}
  	V_\mathrm{c} 					& V_\mathrm{cm}					& V_\mathrm{co} \\	
	V_\mathrm{cm}^{T} 	& V_\mathrm{m}						& V_\mathrm{mo} \\
	V_\mathrm{cm}^{T} 	& V_\mathrm{mo}^{T}	& V_\mathrm{o}
\end{pmatrix}
\end{equation}
where each entry is a $2\times2$ matrix describing the (cross-)correlation of the quadrature operators of the cavity (c), mechanical (m) and filtered output modes (o). 
To quantify the amount of entanglement, we evaluate the logarithmic negativity \cite{Eisert2001, Vidal2002}
\begin{equation}
  E_\mathcal{N}=\max\left\{0,-\log\left(2\sqrt{\frac{\Sigma - \sqrt{\Sigma^2-4 \det(U)}}{2}}\right)\right\}
\end{equation}
with
\begin{equation}
  \Sigma = \det(V_\mathrm{m})+\det(V_\mathrm{o})-2 \det(V_\mathrm{cm}	)
\end{equation}
  and the $4\times4$ correlation matrix of the mechanics and the filtered output mode
\begin{equation}
  U= \begin{pmatrix}	
  	V_\mathrm{m}						& V_\mathrm{mo} \\
	V_\mathrm{mo}^{T}					& V_\mathrm{o}\end{pmatrix}.
\end{equation}

\section{Experimental details}

\subsection{Sample fabrication}
As noted, a centerpiece of our setup is the silicon nitride membrane resonator, embedded in a two-dimensional silicon PnC structure. Upon low-pressure chemical vapor deposition (LPCVD) of stoichiometric silicon nitride onto a 500 micron thick single-crystal silicon wafer, the silicon is etched in potassium hydroxide (KOH), stopping the etch a few micrometers short of releasing the membranes. A PnC structure is etched into the substrate via deep reactive ion etching, before completing the fabrication process with a short KOH etch, thus fully releasing the silicon nitride membranes. More details on the fabrication process can be found in \cite{Tsaturyan2014}.

\begin{figure*}[htb]
\includegraphics[width=0.65\linewidth]{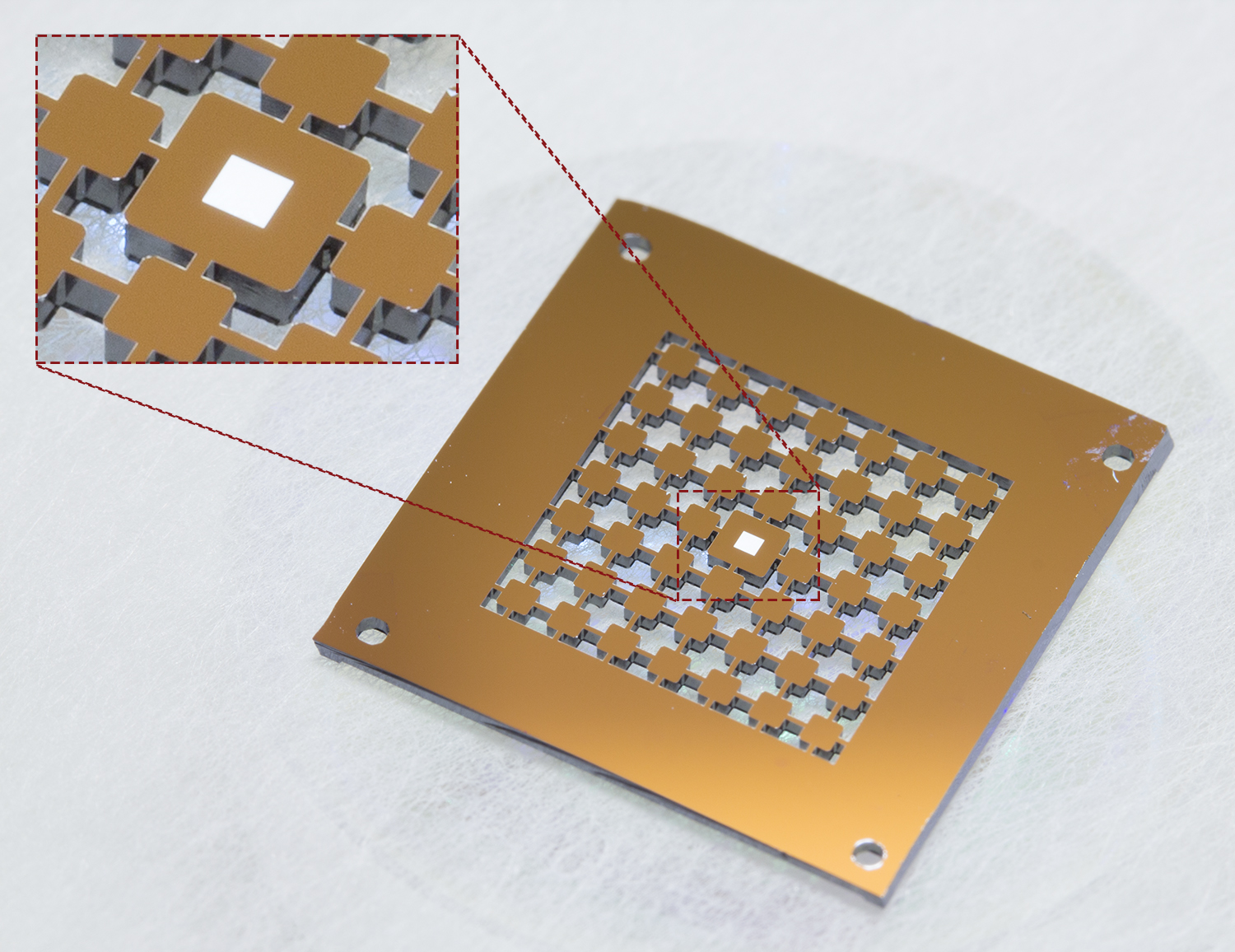}
\caption{Membrane resonator shielded by a two-dimensional phononic crystal structure. The inset shows a close-up of the defect and the membrane resonator.}
\end{figure*}

\subsection{Setup}

The light source used is an MSquared SolsTis laser, delivering single-mode laser light with shot-noise limited phase and amplitude fluctuations for frequencies above 1 MHz. The light is phase-modulated by a fibre-EOM coupled to a mode-cleaning single-mode fibre delivering the light to the cavity. The cavity itself is embedded in a helium flow cryostat. After the cavity, the light is directly detected on a home-built photodetector, with a transmission of the cryostat window in excess of $99\%$, and a detector quantum efficiency of $80\%$.

\subsection{Numerical parameters}

\begin{center}
\begin{tabular}{lcr}
\toprule
Quantity & Symbol & Value  \\
\midrule
Detection efficiency & $\etaDet$ 				& 0{.}80		\\ 
Outcoupler transmission & $\kappa_\mathrm{T}/2\pi$		&13{.}4\,{MHz}		\\
Incoupler transmission & $\kappa_\mathrm{R}/2\pi$		&0{.}6	\,{MHz}		\\
Cavity losses & $\kappa_\mathrm{L}/2\pi$		&0	\,{MHz}		\\
Laser detuning & $\Delta/2\pi$ 						&-1{.}8  \,{MHz}       \\%
Intra-cavity photon number & $n$									& $27\times 10^6$                    \\%
Bare optomechanical coupling & $\vcr/2\pi $   						& 115    \,{Hz}          \\%
Mechanical frequency & $\Om/2\pi$   						& 1928  \,{kHz}	              \\
Mechanical damping rate & $\Gm/2\pi$   						& 170    \,{mHz}             \\
Bath temperature & $T$ & 10\,{K}\\
\bottomrule
\end{tabular}

\end{center}


\subsection{Shot noise calibration}

For the measurements of the squeezing spectra we use a home-built balanced detector with two high-efficiency PIN photodiodes. In calibrating our photodetector we make use of two common techniques, namely calibration via balanced detection and using a thermal light source.
For the former technique the light from the laser is taken directly after the EOM, split and focused onto the two photodiodes. We ensure that the overall power drift is less than a few percent over the acquisition time. The DC photo-current is measured with one of the photodiodes blocked and tuned to a DC-level close to the ones of the recorded squeezing spectra. The second photodiode is unblocked and the difference photo-current is recorded using the same acquisition chain as for the measurements with the optomechanical system.
For comparison the reference measurement is rescaled (typically at the $\%$-level) to match the optical power of the squeezing spectra, whereby a $\sim5\%$ contribution from electronic noise is taken into account.
To verify the result of this measurement we use a second calibration technique, which requires illumination of the relevant photodiode with a thermal white light source. Once again, we ensure that the acquisition-time and the DC photo-current match the ones for the recorded spectrum in Figure 4.

\section*{Localising the beam on the membrane}
\label{sec:local-beam-membr}

In our multimode system, not all membrane modes couple equally strongly to the light. The laser beam illuminates a certain spot on the membrane, where different modes have different displacement amplitudes. We assign to each optomechanical mode, ($i$, $j$), a coupling $G_{ij}$, given by
\begin{equation}
  \label{eq:etaoverlapdefinition1}
  G_{ij} = \eta_{ij}G,
\end{equation}
where $G=\partial \omega_\text{cav}/\partial z_m$ and $\eta_{ij}$ is the so-called \textit{transverse overlap} factor, given by
\begin{equation}
  \label{eq:etaoverlapdefinition1}
  \eta_{ij}(x, y) = \int_\mathcal{D} \dd x' \dd y'\ \sin(ik_x x') \sin(jk_y y') I(x',y', x, y),
\end{equation}
where $\mathcal{D}$ is the domain of the membrane and $I(x', y', x, y)$ is the normalised intensity profile of a laser beam centered at $(x, y)$ . 
To our ends, we exclusively work with the TEM$_{00}$ cavity mode, in which case the integral can be computed analytically to yield 
\begin{equation}
  \label{eq:etaoverlapresult}
  \eta_{ij}(x, y) = \exp\left[-\frac{w^2(z_m)}{8}\left(i^2 k^2_x + j^2 k_y^2\right)\right] \sin(ik_xx) \sin(jk_yy),
\end{equation}
where $w(z_m)$ is the beam width at the membrane, in our case 39 $\upmu$m.

From measurements of different coupling rates, we may then estimate the beam position on the membrane by comparing measured transverse overlaps to the model prediction of equation \eqref{eq:etaoverlapresult}. We measure mode couplings via fitting the OMIT response (see main text) of each mode. This yields the cavity-enhanced coupling $g_{ij}$, given by
\begin{equation}
  \label{eq:transverse0omitfitg}
  g_{ij}=B \frac{\eta_{ij}}{\sqrt{\Omega_m^{(i,j)}}},
\end{equation}
where the factor $B$ is given by
\begin{equation}
  \label{eq:transverseO-B}
  B = \sqrt{\frac{\hbar \bar{n}_\text{cav}}{2m_\text{eff}}}G,
\end{equation}
and is common for all mechanical modes and the mode frequency, $\Omega^{(i, j)}_m$, is given by
\begin{equation}
  \label{eq:Appendix-modefrequency_ij}
  \Omega^{(i, j)}_m = \pi\sqrt{\frac{\mathcal{T}}{\rho}}\sqrt{\frac{i^2}{L_x^2}+\frac{j^2}{L_y^2}},
\end{equation}
where $\mathcal{T}$ is the tensile stress of the membrane and $\rho$ is the density.
For a fixed number, $N$, of measured mode couplings, we construct a unit vector $\vec{v}_\text{data}$ as
\begin{equation}
  \label{eq:chap3-transverse0-datavec}
 \vec{v}_\text{data} = \mathcal{N}[\eta_{i_1j_1}, \eta_{i_2j_2}, \ldots]^T, 
\end{equation}
by multiplying out the mode frequency dependence  \eqref{eq:transverse0omitfitg}. For a given set of ($x$,~$y$)-points (a grid spanning the membrane), we may then, for each point, form a unit vector with the model prediction from equation \eqref{eq:etaoverlapresult},
\begin{equation}
  \label{eq:chap3-transverseO-modelvec}
   \vec{v}_\text{model}(y, x) = \mathcal{N}(x, y)[\eta_{i_1j_1}(x,y), \eta_{i_2j_2}(x,y), \ldots]^T,
\end{equation}
where $\mathcal{N}(x, y)$ is the position-dependent normalisation factor.
If we define an error vector, $\vec{e}$, as
\begin{equation}
  \label{eq:chap3-transverseO-errorvec}
  \vec{e}(x, y) = \vec{v}_\text{data} - \vec{v}_\text{model}(x, y),
\end{equation}
then the most likely position, $(x_0, y_0)$, is the position minimising $\chi^2(x, y)$, where
\begin{equation}
  \label{eq:chap3-chisquaredef}
  \chi^2(x, y) = \sum_{\ell=1}^N e_\ell^2(x, y).
\end{equation}
The estimated uncertainty is then (the number $2$ is the number of model parameters, i.e.\ $x$ and $y$)
\begin{equation}
  \label{eq:chap3-chisquaresigma}
  \sigma^2=\chi^2(x_0,y_0)/(N-2)
\end{equation}
and finally the likelihood function of where the beam is positioned is given by
\begin{equation}
  \label{eq:chap3-likelihoodfunction}
  L(x, y)=\frac{1}{2\pi\sigma^2}\prod_{\ell=1}^N \exp\left( -\frac{ e^2_\ell(x, y)}{2\sigma^2}\right).
\end{equation}
\begin{figure}[h]
  \centering
  \includegraphics[width=0.75\textwidth]{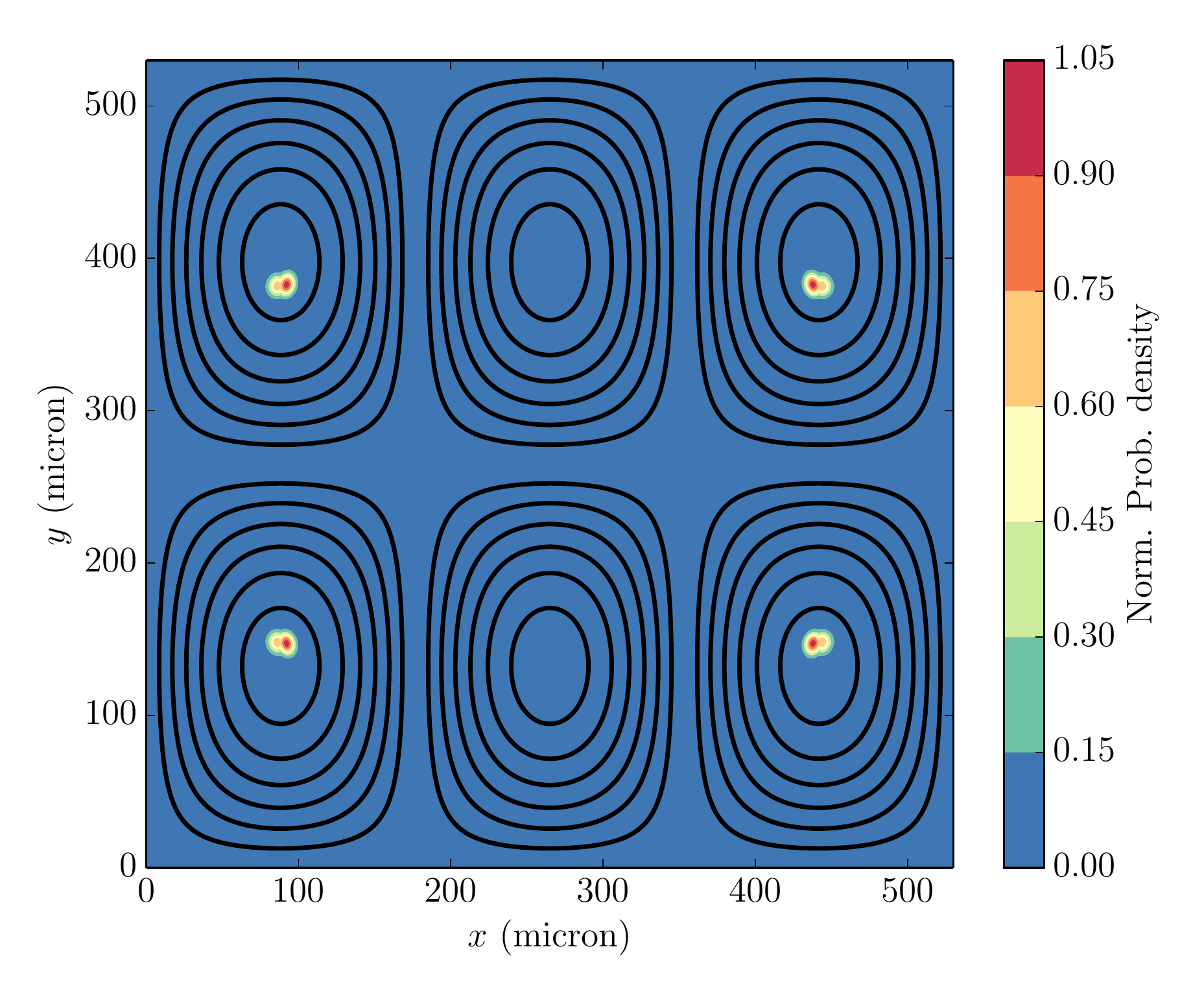}
  \caption{Beam position estimate from $N=36$ modes. Also shown are the mode contours of the ($3$, $2$)-mode aimed for.}
  \label{fig:localisation}
\end{figure}

\end{document}